\documentclass[longauth]{aa}
\usepackage{graphicx}
\usepackage{txfonts}
\usepackage{xcolor}
\usepackage{hyperref}
\hypersetup{colorlinks, linkcolor=blue, citecolor=blue, urlcolor=blue}
\usepackage{amsmath}
\usepackage{ulem}
\usepackage{soul}
\usepackage{enumitem}% http://ctan.org/pkg/enumitem
\newcommand{\muvec}{\mbox{\boldmath $\mu$}}

\newcommand{\te}{t_{\rm E}}
\newcommand{\thetae}{\theta_{\rm E}}

\newcommand{\pie}{\pi_{\rm E}}

\newcommand{\dl}{D_{\rm L}}
\newcommand{\ds}{D_{\rm S}}

\definecolor{brown}{rgb}{0.59, 0.29, 0.0}
\definecolor{darkgreen}{rgb}{0.0, 0.42, 0.24}
\definecolor{darkblue}{rgb}{0.01, 0.31, 0.59}
\definecolor{darkblue}{rgb}{0.0, 0.25, 0.42}
\definecolor{blue}{rgb}{0.0,0.0,1.0}
\definecolor{green}{rgb}{0.0,1.0,0.0}

\begin{document}

%\title{Multi-planet microlensing system OGLE-2019-BLG-0468L composed of two giant planets orbiting around a G-type star}
\title{OGLE-2023-BLG-0836L: The sixth microlensing planet in a binary stellar system}
\titlerunning{OGLE-2023-BLG-0836L: Microlensing planet in binary stellar system}

\author{
% leading author -----------------------------
     Cheongho~Han\inst{\ref{inst1}} 
\and Andrzej~Udalski\inst{\ref{inst2}} 
\and Youn~Kil~Jung\inst{\ref{inst3}}  
\and Andrew~Gould\inst{\ref{inst4}}\inst{\ref{inst5}}              
\and Doeon~Kim\inst{\ref{inst1}} 
\\
(Leading authors)
\\
% KMTNet ---------------------------
     Michael D. Albrow\inst{\ref{inst6}} 
\and Sun-Ju~Chung\inst{\ref{inst3}}  
\and Kyu-Ha~Hwang\inst{\ref{inst3}}  
\and Chung-Uk~Lee\inst{\ref{inst3}}  
\and Yoon-Hyun~Ryu\inst{\ref{inst3}}  
\and Yossi~Shvartzvald\inst{\ref{inst7}}    
\and In-Gu~Shin\inst{\ref{inst8}}  
\and Jennifer~C.~Yee\inst{\ref{inst8}}   
\and Hongjing~Yang\inst{\ref{inst9}}   
\and Weicheng~Zang\inst{\ref{inst8}}\inst{\ref{inst9}}    
\and Sang-Mok~Cha\inst{\ref{inst3}}\inst{\ref{inst10}} 
\and Dong-Jin~Kim\inst{\ref{inst3}}
\and Seung-Lee~Kim\inst{\ref{inst3}}\inst{\ref{inst11}}
\and Dong-Joo~Lee\inst{\ref{inst3}}
\and Yongseok~Lee\inst{\ref{inst3}}\inst{\ref{inst10}}
\and Byeong-Gon~Park\inst{\ref{inst3}}
\and Richard~W.~Pogge\inst{\ref{inst5}}
\\
(The KMTNet Collaboration)
\\
% OGLE ---------------------------
     Przemek~Mr{\'o}z\inst{\ref{inst2}}
\and Mateusz~J.~Mr{\'o}z\inst{\ref{inst2}}
\and Micha{\l}~K.~Szyma{\'n}ski\inst{\ref{inst2}}
\and Jan~Skowron\inst{\ref{inst2}}
\and Rados{\l}aw~Poleski\inst{\ref{inst2}}
\and Igor~Soszy{\'n}ski\inst{\ref{inst2}}
\and Pawe{\l}~Pietrukowicz\inst{\ref{inst2}}
\and Szymon~Koz{\l}owski\inst{\ref{inst2}}
\and Krzysztof~A.~Rybicki\inst{\ref{inst2}}\inst{\ref{inst7}}
\and Patryk~Iwanek\inst{\ref{inst2}}
\and Krzysztof~Ulaczyk\inst{\ref{inst12}}
\and Marcin~Wrona\inst{\ref{inst2}}
\and Mariusz~Gromadzki\inst{\ref{inst2}}
\\
(The OGLE Collaboration)
}

\institute{
     Department of Physics, Chungbuk National University, Cheongju 28644, Republic of Korea  \\ \email{cheongho@astroph.chungbuk.ac.kr} \label{inst1}    % (1)
\and Astronomical Observatory, University of Warsaw, Al.~Ujazdowskie 4, 00-478 Warszawa, Poland   \label{inst2}                                          % (2) 
\and Korea Astronomy and Space Science Institute, Daejon 34055, Republic of Korea    \label{inst3}                                                       % (3)
\and Max Planck Institute for Astronomy, K\"onigstuhl 17, D-69117 Heidelberg, Germany   \label{inst4}                                                    % (4)
\and Department of Astronomy, The Ohio State University, 140 W. 18th Ave., Columbus, OH 43210, USA  \label{inst5}                                        % (5)
\and University of Canterbury, Department of Physics and Astronomy, Private Bag 4800, Christchurch 8020, New Zealand  \label{inst6}                      % (6)
\and Department of Particle Physics and Astrophysics, Weizmann Institute of Science, Rehovot 76100, Israel \label{inst7}                                 % (7)
\and Center for Astrophysics $|$ Harvard \& Smithsonian 60 Garden St., Cambridge, MA 02138, USA  \label{inst8}                                           % (8)
\and Department of Astronomy and Tsinghua Centre for Astrophysics, Tsinghua University, Beijing 100084, China   \label{inst9}                            % (9)
\and School of Space Research, Kyung Hee University, Yongin, Kyeonggi 17104, Republic of Korea  \label{inst10}                                           % (10)  
\and Korea University of Science and Technology, 217 Gajeong-ro, Yuseong-gu, Daejeon, 34113, Republic of Korea \label{inst11}                            % (11)
\and  Department of Physics, University of Warwick, Gibbet Hill Road, Coventry, CV4 7AL, UK\label{inst12}                                                % (12)
}
\date{Received ; accepted}

% \abstract{}{}{}{}{} 
% 5 {} token are mandatory
\abstract
% context heading (optional)
% {} leave it empty if necessary  
{}
% aims heading (mandatory)
{
Light curves of microlensing events occasionally deviate from the smooth and symmetric
form of a single-lens single-source event. While most of these anomalous events can be 
accounted for by employing a binary-lens single-source (2L1S) or a single-lens binary-source 
(1L2S) framework, it is established that a small fraction of events remain unexplained by 
either of these interpretations. We carry out a project in which  data collected by high-cadence  
microlensing surveys were reinvestigated with the aim of uncovering the nature of anomalous 
lensing events with no proposed 2L1S or 1L2S models.
}
% methods heading (mandatory)
{
From the project, we find that the anomaly appearing in the lensing event OGLE-2023-BLG-0836 
cannot be explained by the usual interpretations and conduct a comprehensive analysis of the 
event. From thorough modeling of the light curve under sophisticated lens-system configurations, 
we have arrived at the conclusion that a triple-mass lens system is imperative to account for 
the anomaly features observed in the lensing light curve.
}
% results heading (mandatory)
{
From the Bayesian analysis using the measured observables of the event time scale and angular 
Einstein radius, we determine that the least massive component of the lens has a planetary 
mass of $4.36^{+2.35}_{-2.18}~M_{\rm J}$. This planet orbits within a stellar binary system 
composed of two stars with masses $0.71^{+0.38}_{-0.36}~M_\odot$ and $0.56^{+0.30}_{-0.28}~M_\odot$. 
This lensing event signifies the sixth occurrence of a planetary microlensing system in which 
a planet belongs to a stellar binary system.
}
% conclusions heading (optional), leave it empty if necessary 
{}


\maketitle

\section{Introduction}\label{sec:one}

The light curve of a microlensing event involving a single lens and a single source (1L1S) is
represented by
\begin{equation}
F(t) = A(t) F_s + F_b;\qquad A(t) = { u^2+2\over u(u^2+4)^{1/2}}, 
\label{eq1}
\end{equation}
where $A(t)$ is the lensing magnification, $F_s$ and $F_b$ denote the respective flux values 
originating from the source and blended light components, and $u$ represents the projected 
lens-source separation normalized to the angular Einstein radius $\thetae$. The lensing 
magnification varies in time as the lens-source separation changes due to their relative 
motion as 
\begin{equation}
u(t) = \left[ u_0^2 + { (t-t_0)^2 \over \te} \right]^{1/2},
\label{eq2}
\end{equation}
where $u_0$ and $t_0$ represent the minimum lens-source separation (scaled to $\thetae$) and 
the corresponding time, and $\te$ is the Einstein time scale. The resulting light curve is 
characterized by a smooth and symmetric form \citep{Paczynski1986}.

Light curves in microlensing events occasionally exhibit deviations from the standard 1L1S 
form.  These deviations are, in most cases, attributed to two primary factors: the potential 
binarity of the lens, as described by \citet{Mao1991}, and the binarity of the source, as 
noted by \citet{Griest1992}. In the case of a binary-lens (2L1S) event, the lensing system 
creates a complex pattern of caustics. These caustics represent specific positions of the 
source at which the lensing magnification for a point source diverges to infinity. When a 
source crosses the caustic, a pair of new images are created or disappear, resulting in a 
complicated lensing light curve that deviates from the 1L1S form. In the case of a binary-source 
(1L2S) event, the lensing magnification corresponds to the mean of the magnifications associated 
with the individual binary source stars, $A_1$ and $A_2$, weighted by the flux contributions of 
the component source stars, $F_1$ and $F_2$, that is, 
\begin{equation}
A = {A_1 F_1 +A_2 F_2  \over F_1 + F_2}.
\label{eq3}
\end{equation}
As a consequence, the light curve of a 1L2S event displays deviations from the standard 1L1S
form.

Since 2016, the Korea Microlensing Telescope Network (KMTNet) has been conducting a 
microlensing survey with frequent observations of stars located in the direction of the 
Galactic bulge \citep{Kim2016}. Among about 3000 microlensing events that are being annually 
detected from the survey, about 10\% of events exhibit anomalies in the lensing light curves. 
While the majority of these anomalous lensing events can be explained by applying a 2L1S or 
a 1L2S framework, it is known that a small fraction of events defy explanation under either 
of these interpretations. The challenge in precisely characterizing the peculiarities of 
these events hints at the necessity for more advanced models to interpret the observed 
anomalies.

We have conducted a project that involved revisiting microlensing data collected by the 
KMTNet survey. The primary goal of this project is to uncover instances of anomalous lensing 
events for which the conventional 2L1S or 1L2S models had not been previously proposed. 
Through this investigation, we have identified multiple occurrences that required the 
application of advanced modeling approaches beyond the standard 2L1S or 1L2S frameworks. 
\citet{Han2019} found that the anomaly appearing in the light curve of the lensing event 
OGLE-2018-BLG-1011, which corresponds to KMTNet event KMT-2018-BLG-2122, could be explained 
with a triple-lens (3L1S) model in which the lens is composed of two giant planets and their 
host star. Through a detailed analysis of the light curve for the lensing event OGLE-2018-BLG-1700 
(KMT-2018-BLG-2330), \citet{Han2020a} identified the triple nature of the lens by decomposing 
the anomaly into two parts produced by two binary-lens pairs. In one of these binary pairs, 
the mass ratio between the lens components is approximately $q \sim 0.01$, while in the other 
pair, the mass ratio is around $\sim 0.3$, suggesting that lens is a planetary system in a 
binary. Through a careful examination of the central anomaly observed in the lensing curve 
of the highly magnified event KMT-2019-BLG-1953, \citet{Han2020b} found that the discrepancies 
from the 2L1S model were significantly reduced when an additional planetary lens companion or 
a source companion was introduced, although distinguishing these two interpretations was 
difficult within the precision of the photometric data.  In their study, \citet{Han2021a} 
determined that the anomalous characteristics observed in the lensing light curve of the 
event KMT-2019-BLG-0797 could be accounted for by a 2L2S model, in which both the lens and 
source are binary systems. By analyzing the event KMT-2019-BLG-1715, for which the lensing 
light curve exhibited two short-term deviation features from a caustic-crossing 2L1S light 
curve, \citet{Han2021b} suggested a five-body (lens+source) model, in which one deviation 
feature was generated by a planetary-mass third body of the lens, and the other feature was 
generated by a faint source companion, and thus the event is a very complex five-body system 
composed of three lens masses (planet + two stars) and two source stars.  \citet{Han2021c} 
found that KMT-2018-BLG-1743 is another planetary lensing event occurring on two source 
stars.  In their analysis of the anomalies observed in the lensing event OGLE-2019-BLG-0304 
(KMT-2019-BLG-2583), \citet{Han2021d} put forward two rivaling models.  The first 3L1S model 
suggests the presence of a planetary-mass third body situated near the primary lens of the 
binary lens system. The second 2L2S model proposes the existence of an additional nearby 
companion to the source. \citet{Zang2021} found that the central anomaly in the lensing light 
curve of the event KMT-2020-BLG-0414 was produced by a triple-lens system, which consists of 
an Earth-mass planet and its binary host. Through the investigation of the anomalous lensing 
event KMT-2021-BLG-1077, \citet{Han2022a} identified that the lens of the event is a 
multi-planetary system in which two giant planets orbit a very low-mass star. It was found 
by \citet{Han2022b} that the dual bump anomaly feature in the light curve of the lensing 
event KMT-2021-BLG-1898 could be explained with a 2L2S model, in which the lens contains 
a giant planet and the source is a binary composed of a turnoff star and a K-type dwarf.  
\citet{Han2022c} found that the planetary signal in the lensing light curve of the event 
KMT-2021-BLG-0240 was deformed either by an extra planetary lens component or by a companion 
to the source, although the 3L1S and 2L2S interpretations could not be distinguished with 
the available data. \citet{Han2023a} found that the lensing events OGLE-2018-BLG-0584 and 
KMT-2018-BLG-2119 were generated by 2L2S lens systems. From the analysis of the lensing 
event KMT-2021-BLG-1122, \citet{Han2023b} revealed that the anomaly appeared in the light 
curve was produced by a 3L1S system, which consists of three stars.

In this study, we provide a comprehensive analysis of the microlensing event
OGLE-2023-BLG-0836/KMT-2023-BLG-1144, for which no existing model has successfully 
explained the anomaly observed in the lensing light curve. The anomaly in question 
presents two distinctive features: a caustic-crossing pattern and a strong cusp-approaching
 peak. Our investigation has led us to the conclusion that the inclusion of a triple-mass 
lens system is imperative to adequately account for all the anomalous features in the 
lensing light curve of the event.

% Figure 1 ------------------------------------------------------
\begin{figure}[t]
\includegraphics[width=\columnwidth]{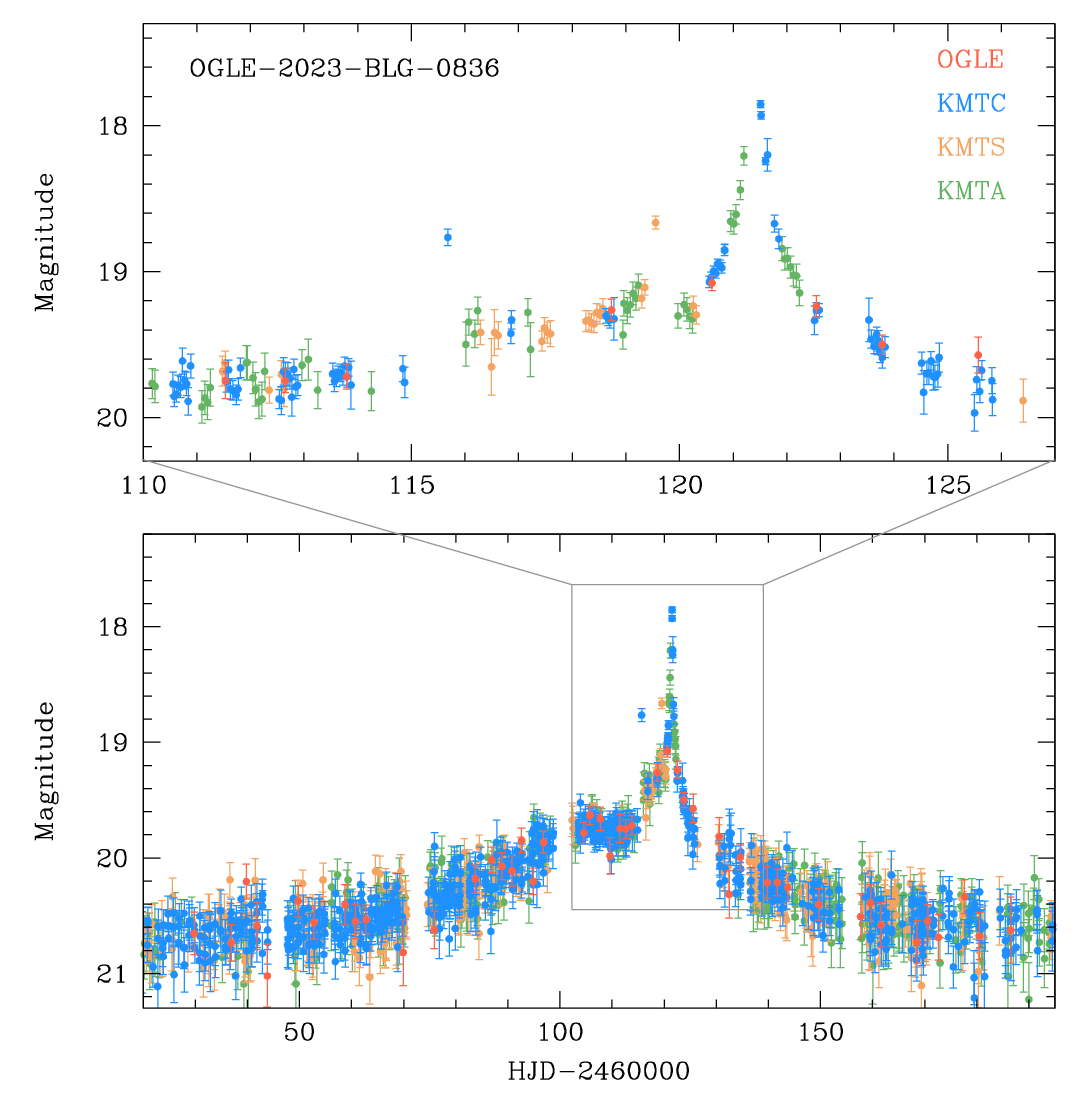}
\caption{
Lensing light curve of OGLE-2023-BLG-0836.  The top panel displays an enlarged view 
focused on the vicinity of the anomaly.
}
\label{fig:one}
\end{figure}
% --------------------------------------------------------------

\section{Observations and data}\label{sec:two}

The source of the microlensing event OGLE-2023-BLG-0836/KMT-2023-BLG-1144 is positioned in
the direction of the Galactic bulge field, with equatorial coordinates $({\rm RA}, 
{\rm Dec})_{\rm J2000} = (17$:48:44.85, $-$23:44:29.47), which correspond to the Galactic 
coordinates $(l, b) = (4^\circ\hskip-2pt .8075, 2^\circ\hskip-2pt .0912)$. In this direction, 
the extinction in the $I$-band is approximately $A_I \sim 2.01$. The magnification of the 
source flux caused by lensing was first detected from the Optical Gravitational Lensing 
Experiment IV \citep[OGLE-IV:][]{Udalski2015} survey on 28 June 2023, corresponding to 
the reduced Heliocentric Julian Date (${\rm HJD}^\prime \equiv {\rm HJD}-2460000= 123$). 
Five days later, the KMTNet group independently found the event and designated it as 
KMT-2023-BLG-1144.  While we initially recognized the anomalous nature of the event through 
a systematic analysis of the KMTNet data gathered during the 2023 season, we have chosen to 
label the event as OGLE-2023-BLG-0836, aligning with the reference ID from the OGLE survey 
that initially detected the event.

% Figure 2 ------------------------------------------------------
\begin{figure}[t]
\includegraphics[width=\columnwidth]{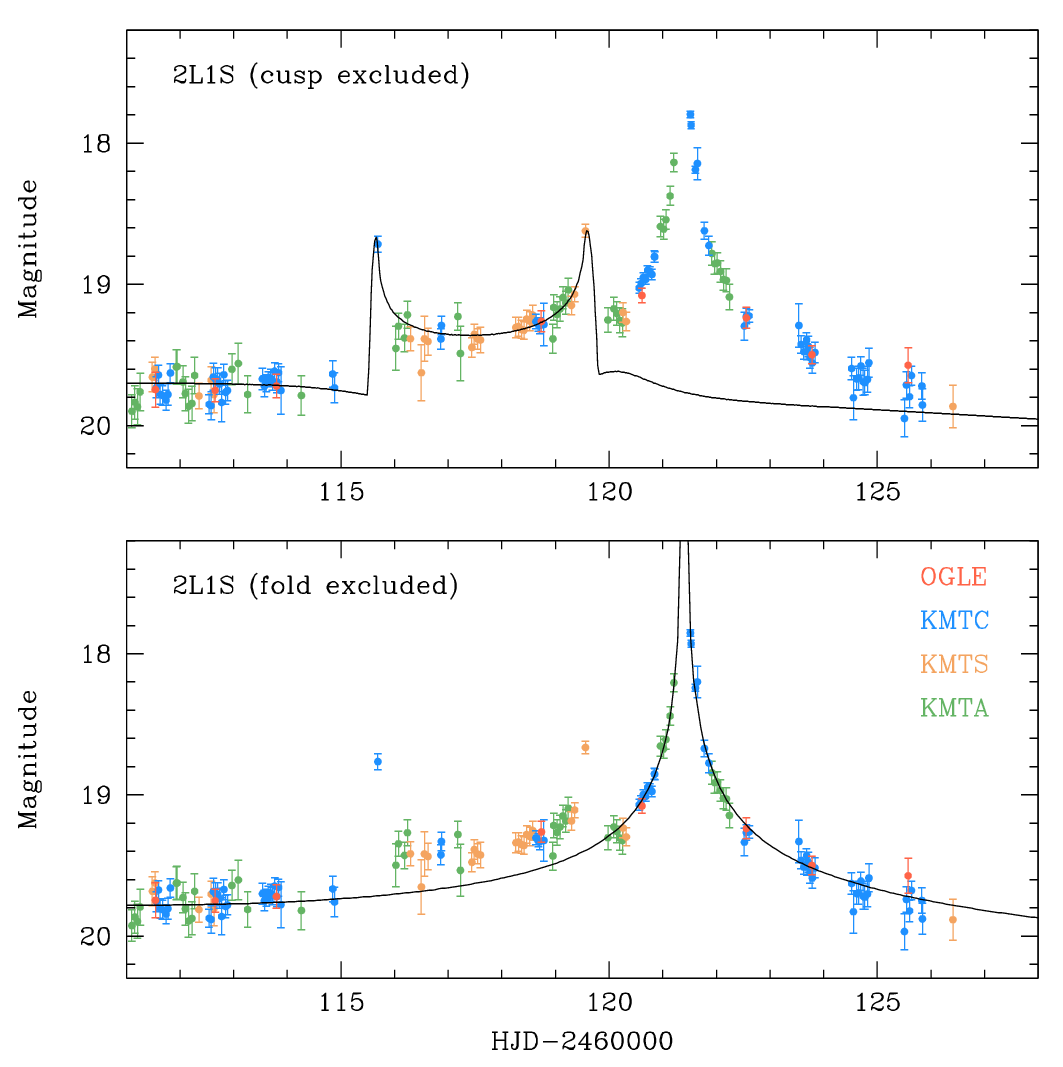}
\caption{
Division of the anomaly into two parts.  The curve presented in the upper panel represents 
a 2L1S model derived by fitting the data excluding those around the cusp-approaching feature, 
while the curve in the lower panel is a 2L1S model obtained by fitting the data with the 
exclusion of the data around caustic-crossing feature. 
}
\label{fig:two}
\end{figure}
% --------------------------------------------------------------

The event observations were conducted using the telescopes operated by individual survey
groups. The KMTNet group employs three identical telescopes, each featuring a 1.6-meter
aperture and a wide-field camera capable of capturing 4 square degrees in a single exposure.
In order to ensure continuous coverage of lensing events, the KMTNet telescopes are
strategically positioned across the three continents in the Southern Hemisphere: Siding
Spring Observatory in Australia (KMTA), Cerro Tololo Interamerican Observatory in Chile
(KMTC), and South African Astronomical Observatory in South Africa (KMTS).  Additionally, 
the OGLE telescope, featuring a 1.3-meter aperture and a camera yielding a 1.4 square 
degree field of view, is located at Las Campanas Observatory in Chile. Images from the 
KMTNet and OGLE surveys were mainly obtained in the $I$ band with the inclusion of some 
$V$-band images taken for source color measurement. Observations of the event by the KMTNet 
and OGLE surveys were done with an hour and about 2~day cadences, respectively. Image 
reduction and photometry for the lensing event were accomplished using automated pipelines 
tailored to the individual surveys, with those developed by \citet{Albrow2017} for the KMTNet 
survey and by \citet{Wozniak2000} for the OGLE survey. For the use of optimal data in the 
analysis, we conducted re-reduction of the KMTNet data using the photometry code developed 
by \citet{Yang2023}.  Additionally, the error bars of the data were adjusted to ensure 
consistency with the data scatter, and to set the $\chi^2$ value per degree of freedom 
(d.o.f.) for each data set to unity following the procedure outlined in \citet{Yee2012}.

% Figure 3 ------------------------------------------------------
\begin{figure}[t]
\includegraphics[width=\columnwidth]{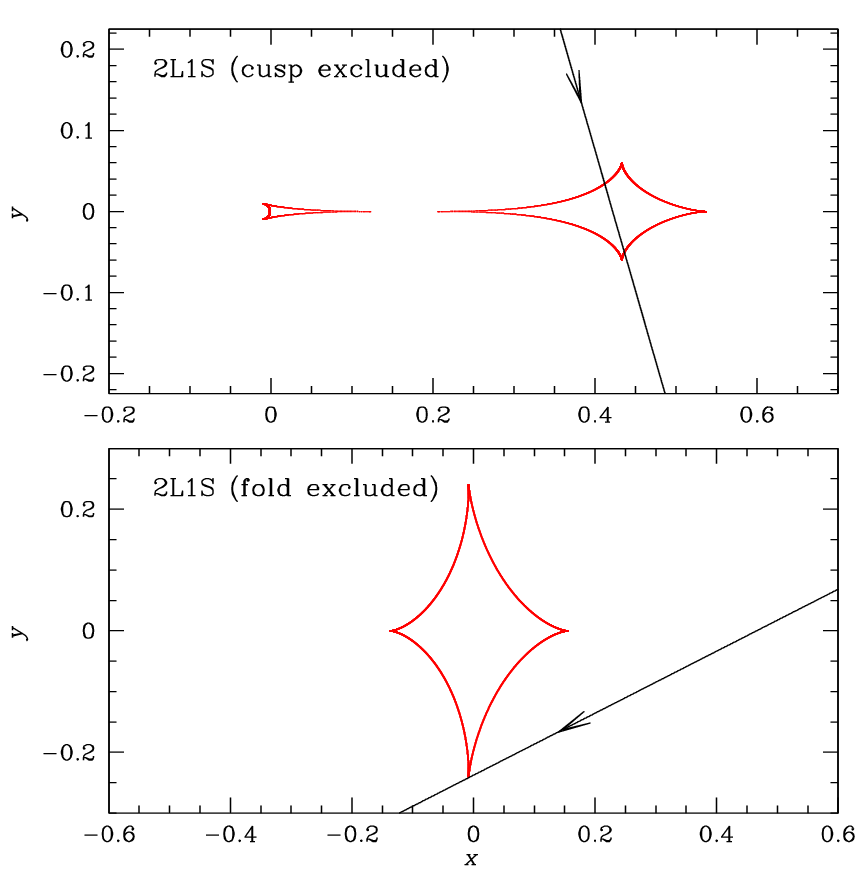}
\caption{
The lens system configurations of the two 2L1S solutions for which the model curves are 
presented in the corresponding panels of Fig.~\ref{fig:two}. In each panel, the red figure 
is the caustic, the line represents the source trajectory, and the arrow on the source 
trajectory indicates the direction of the source motion.  By convention, the abscissa 
of such 2L1S diagrams is defined by the binary axis of the lens system.
}
\label{fig:three}
\end{figure}
% --------------------------------------------------------------

Figure~\ref{fig:one} presents the lensing light curve of OGLE-2023-BLG-0836, revealing 
deviations from the typical smooth and symmetric shape observed in a 1L1S event. These 
deviations are marked by two primary features. The first is the caustic-crossing feature, 
which is characterized by a pair of caustic-crossing spikes at ${\rm HJD}^\prime \sim 115.7$ 
and $\sim 119.6$ and a U-shape trough region between the spikes.  The second feature is the 
strong peak, centered at ${\rm HJD}^\prime \sim 121.3$, which is likely to be produced by 
the source star's approach to a caustic cusp.  The upper panel of Figure~\ref{fig:one} 
offers a detailed view of this anomalous region.  Because caustics in gravitational lensing 
are formed due to the presence of multiple lensing masses, the caustic-crossing feature 
indicates that the lens is comprised of multiple masses.  Furthermore, the approximately 
symmetry between the ascending and descending segments of the peak anomaly feature suggests 
that the feature likely originated from the source's approach to a cusp of a caustic.

% Figure 4 ------------------------------------------------------
\begin{figure*}[t]
\centering
\includegraphics[width=12.0cm]{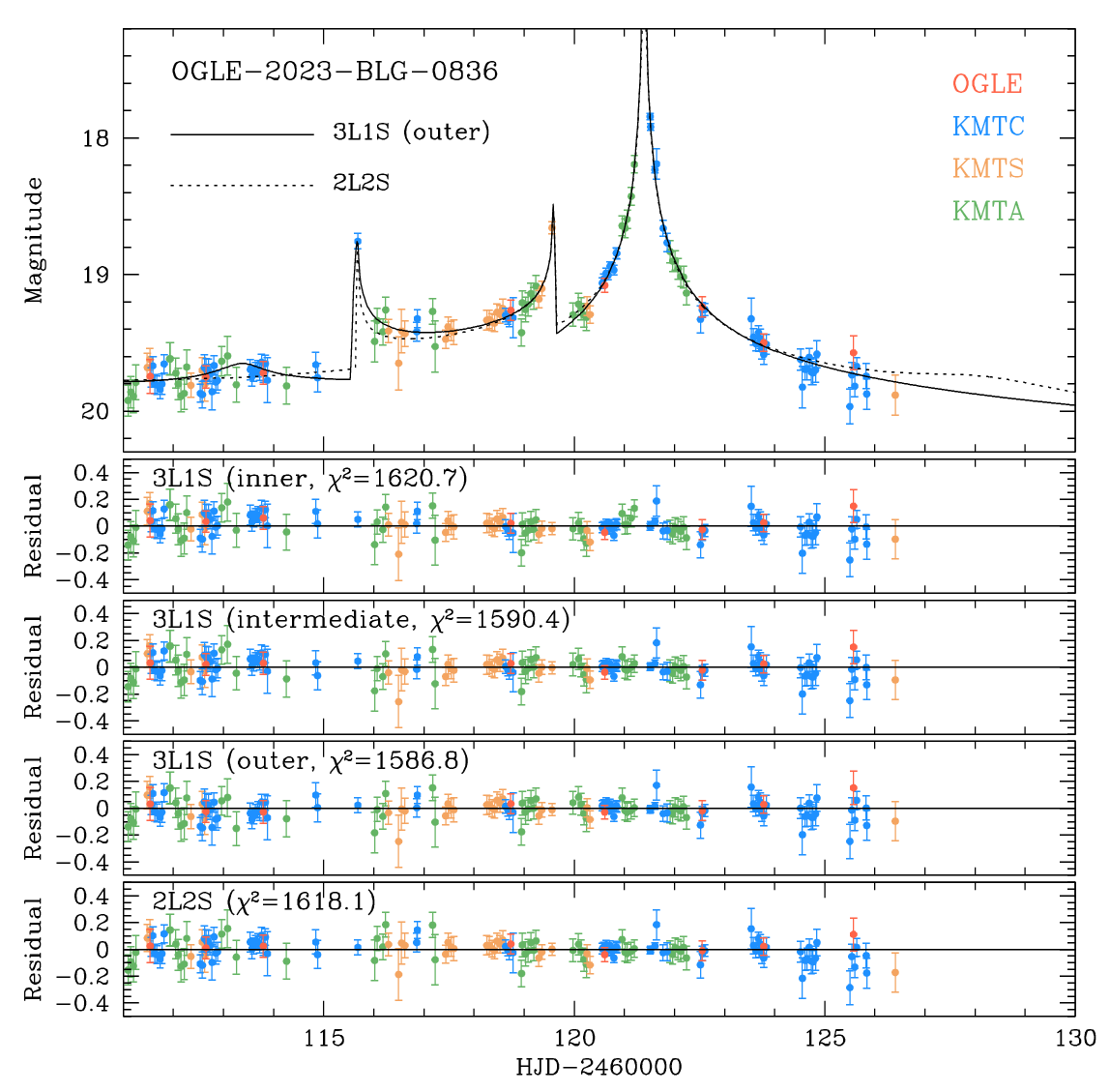}
\caption{
Model curves of the best-fit 3L1S solution (outer solution) and 2L2S solution in the 
region of the anomaly. The lower four panels present the residuals from the three 
degenerate 3L1S solutions (inner, intermediate, and outer solutions) and the 2L2S solution.  
}
\label{fig:four}
\end{figure*}
% --------------------------------------------------------------

\section{Lensing light curve analysis}\label{sec:three}

\subsection{Binary-lens analysis}\label{sec:three-one}

Taking into account the potential involvement of the anomaly features with caustics, 
our analysis commences by modeling the light curve under the interpretation with a 2L1S 
lens-system configuration. This modeling process is conducted to identify a lensing solution, 
which comprises a set of lensing parameters that best describe the characteristics of the 
light curve. Under the approximation of rectilinear relative motion between the lens and 
source, the light curve of a 2L1S event is defined by seven fundamental lensing parameters. 
The initial three parameters $(t_0, u_0, \te)$ characterize the approach of the source to 
the lens. The following three parameters $(s, q, \alpha)$ provide information about the 
binary lens system, and $s$ denotes the projected separation (scaled to $\thetae$) between 
the lens components $M_1$ and $M_2$, $q = M_2/M_1$ represents the mass ratio between these 
lens components, and $\alpha$ is the source trajectory angle, defined as the angle between 
the direction of the relative lens-source proper motion vector $\muvec$ and the $M_1$--$M_2$, 
axis. The last parameter, $\rho$, is defined as the ratio of the angular source radius 
$\theta_*$ to the Einstein radius (normalized source radius), that is, $\rho = \theta_*/
\thetae$, and it characterizes how finite source effects contribute to the deformation of 
the lensing light curve during caustic crossings.

% Table 1 ------------------------------------------------
\begin{table}[t]
%\footnotesize
%\small
%\centering
\caption{Lensing parameters of 2L1S solutions.\label{table:one}}
%\begin{tabular*}{lllllll}
\begin{tabular*}{\columnwidth}{@{\extracolsep{\fill}}lllcc}
\hline\hline
\multicolumn{1}{c}{Parameter}          &
\multicolumn{1}{c}{Cusp excluded}  &
\multicolumn{1}{c}{Fold excluded}    \\
\hline
 $t_0$ (${\rm HJD}^\prime$)    &   $112.007 \pm 0.239$                    &   $114.745 \pm 0.263$     \\ 
 $u_0$                         &   $  0.398 \pm 0.007$                    &   $  0.222 \pm 0.011$     \\ 
 $\te$ (days)                  &   $ 46.34  \pm 0.85 $                    &   $ 53.69  \pm 2.22 $     \\ 
 $s  $                         &   $  1.234 \pm 0.005$                    &   $  0.599 \pm 0.010$     \\ 
 $q  $                         &   $ (3.39  \pm 0.20) \times 10^{-3} $    &   $  0.845 \pm 0.092$     \\ 
 $\alpha$ (rad)                &   $  4.437 \pm 0.012$                    &   $  1.406 \pm 0.021$     \\ 
 $\rho$ ($10^{-3}$)            &   $  1.79  \pm 0.20 $                    &   --                      \\ 
\hline
\end{tabular*}
\tablefoot{ ${\rm HJD}^\prime = {\rm HJD}- 2460000$.  }
\end{table}
% --------------------------------------------------------

In our pursuit of the lensing solution, we categorized the lensing parameters into two
groups. Within the first group, which pertained to the binary parameters $(s, q, \alpha)$, 
we conducted a grid-based exploration to determine the values of parameters $s$ and $q$ 
with multiple initial values of $\alpha$. The other parameters of the second group were 
determined by minimizing $\chi^2$ through the use of the Markov Chain Monte Carlo (MCMC) 
method, which employs an adaptive step size Gaussian sampler, as described in \citet{Doran2004}. 
To assess the presence of degenerate solutions, we examined the $\Delta\chi^2$ map in the 
$s$--$q$ parameter space derived from the grid search. Having identified local solutions, 
we further refined the lensing parameters of the solutions using a downhill approach.  In 
cases for which the discrepancies in $\chi^2$ values among these local solutions were marginal, 
we presented all of them and investigated the causes of degeneracies.

Despite our comprehensive exploration of the parameter space, we were unable to identify 
a 2L1S solution capable of sufficiently explaining both the caustic-crossing and  
cusp-approaching features within the anomaly. This underscores the necessity of employing 
a more sophisticated model for a comprehensive understanding of the observed anomaly 
features.

% Table 2 ------------------------------------------------
\begin{table*}[t]
%\footnotesize
%\small
%\centering
\caption{Lensing parameters of 3L1S solutions.\label{table:two}}
\begin{tabular}{lllllll}
%\begin{tabular}{\columnwidth}{@{\extracolsep{\fill}}lllcc}
\hline\hline
\multicolumn{1}{c}{Parameter}     &
\multicolumn{1}{c}{Inner}   &
\multicolumn{1}{c}{Intermediate}    &
\multicolumn{1}{c}{Outer}    \\
\hline
 $\chi^2$/d.o.f.              &  $1620.8/1619        $  &   $1591.0/1619         $   &   $1586.8/1619         $  &    \\
 $t_0$ (${\rm HJD}^\prime$)   &  $60115.37 \pm 0.17  $  &   $60114.70 \pm 0.22   $   &   $60115.04 \pm 0.20   $  &    \\
 $u_0$                        &  $0.1773 \pm 0.0073  $  &   $0.2270 \pm 0.0080   $   &   $0.1886 \pm 0.0057   $  &    \\
 $\te$ (days)                 &  $58.98 \pm 1.76     $  &   $53.01 \pm 1.30      $   &   $60.85 \pm 1.60      $  &    \\
 $s_2$                        &  $0.5717 \pm 0.0082  $  &   $0.6022 \pm 0.0065   $   &   $0.5661 \pm 0.0057   $  &    \\
 $q_2$                        &  $0.593 \pm 0.053    $  &   $0.882 \pm 0.068     $   &   $0.793 \pm 0.071     $  &    \\
 $\alpha$ (rad)               &  $-0.439 \pm 0.013   $  &   $-0.485 \pm 0.016    $   &   $-0.467 \pm 0.016    $  &    \\
 $s_3$                        &  $1.1344 \pm 0.0073  $  &   $1.1557 \pm 0.0096   $   &   $1.1136 \pm 0.0066   $  &    \\
 $q_3$ ($10^{-3}$)            &  $4.88 \pm 0.82      $  &   $4.97 \pm 0.45       $   &   $5.86 \pm 0.48       $  &    \\
 $\psi$ (rad)                 &  $5.081 \pm 0.011    $  &   $5.100 \pm 0.013     $   &   $5.079 \pm 0.012     $  &    \\
 $\rho$ ($10^{-3}$)           &  $0.60 \pm 0.19      $  &   $0.60 \pm 0.19       $   &   $0.60 \pm 0.19       $  &    \\
\hline
\end{tabular}
%\tablefoot{ ${\rm HJD}^\prime = {\rm HJD}- 2460000$.  }
\end{table*}
% --------------------------------------------------------

% Figure 5 ------------------------------------------------------
\begin{figure}[t]
\includegraphics[width=\columnwidth]{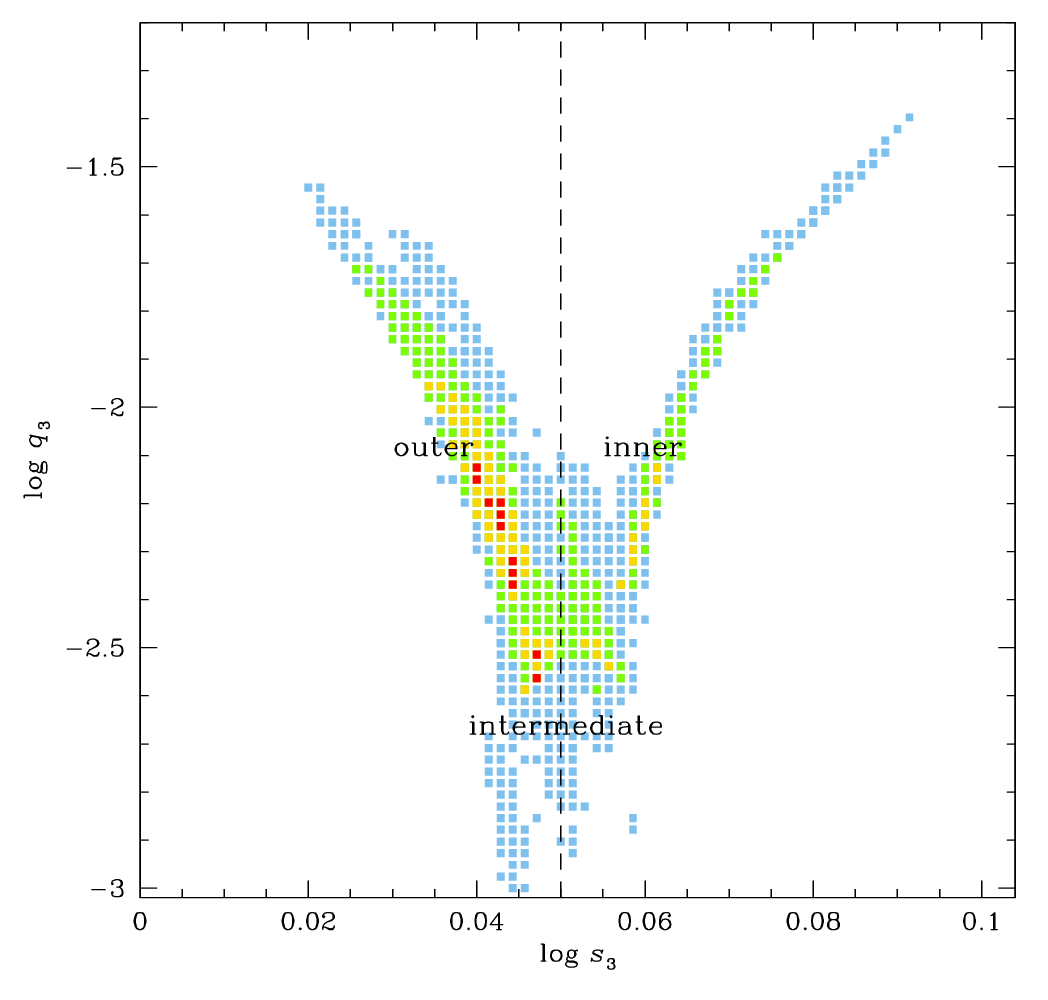}
\caption{
$\Delta\chi^2$ map on the $(\log s_3, \log q_3)$ parameter plane. The color-coding is 
configured to correspond to data points based on their $\Delta\chi^2$ values: 
red    for $\Delta\chi^2 \leq 1^2 n$, 
yellow for $ \leq 2^2 n $,
green  for $ \leq 3^2 n $, and 
cyan   for $ \leq 4^2 n $, where $n=4$. The three distinct local solutions are marked as 
"inner", "intermediate", and "outer". The dashed vertical line represents the  geometric 
mean $(s_{3,{\rm in}}\times s_{3,{\rm out}})^{1/2}$ of the planetary separations for the 
inner and outer solutions.
}
\label{fig:five}
\end{figure}
% --------------------------------------------------------------

While the 2L1S model cannot simultaneously accommodate both the caustic-crossing and
cusp-approaching features, we have identified that each anomaly feature can be adequately
approximated by a 2L1S model. This is illustrated in Figure~\ref{fig:two}, in which we present 
two 2L1S model curves depicting the individual anomaly features. The curve presented in the 
upper panel represents a 2L1S model derived by fitting the data excluding the observations 
around the cusp-approaching feature during the time interval $120 \lesssim {\rm HJD}^\prime 
\lesssim 125$, while the curve shown in the lower panel is a model obtained by fitting the 
data with the exclusion of the data around the caustic feature within the time range $114 
\lesssim {\rm HJD}^\prime \lesssim 120$. See also Figure 2 of \citet{Han2020a} for another 
example of an anomaly that can be divided into two parts produced by two 2L1S events.  We 
found a solution with binary parameters $(s, q) \sim (0.60, 0.85)$ from the 2L1S fit to 
the data excluding the caustic-crossing feature. Similarly, we also identified a solution 
with $(s, q) \sim  (1.23, 3.4\times 10^{-3})$ from the fit to the data excluding those 
around the cusp-approaching feature.  The full lensing parameters of the pair of 2L1S 
solutions are presented in Table~\ref{table:one}.  Regarding the lensing parameters in 
the model describing the caustic-crossing feature, we note that the mass ratio between 
the lens components is on the order of $10^{-3}$, indicating that the companion is very 
likely to be a planet-mass object.

Figure~\ref{fig:three} displays the lens-system configurations corresponding to the individual 
2L1S models presented in Figure~\ref{fig:two}. In each panel, the red figure comprising concave 
closed curves represents the caustic, and the arrowed line represents the trajectory of the 
source. The configuration illustrates that the caustic-crossing feature resulted from the source 
star's crossing over the planetary caustic induced by a planetary companion. On the other hand, 
the peak feature of the anomaly was produced by the source star's close approach to the sharp 
off-axis cusp of a caustic induced by a binary lens composed of roughly equal masses.

% Figure 6 ------------------------------------------------------
\begin{figure}[t]
\includegraphics[width=\columnwidth]{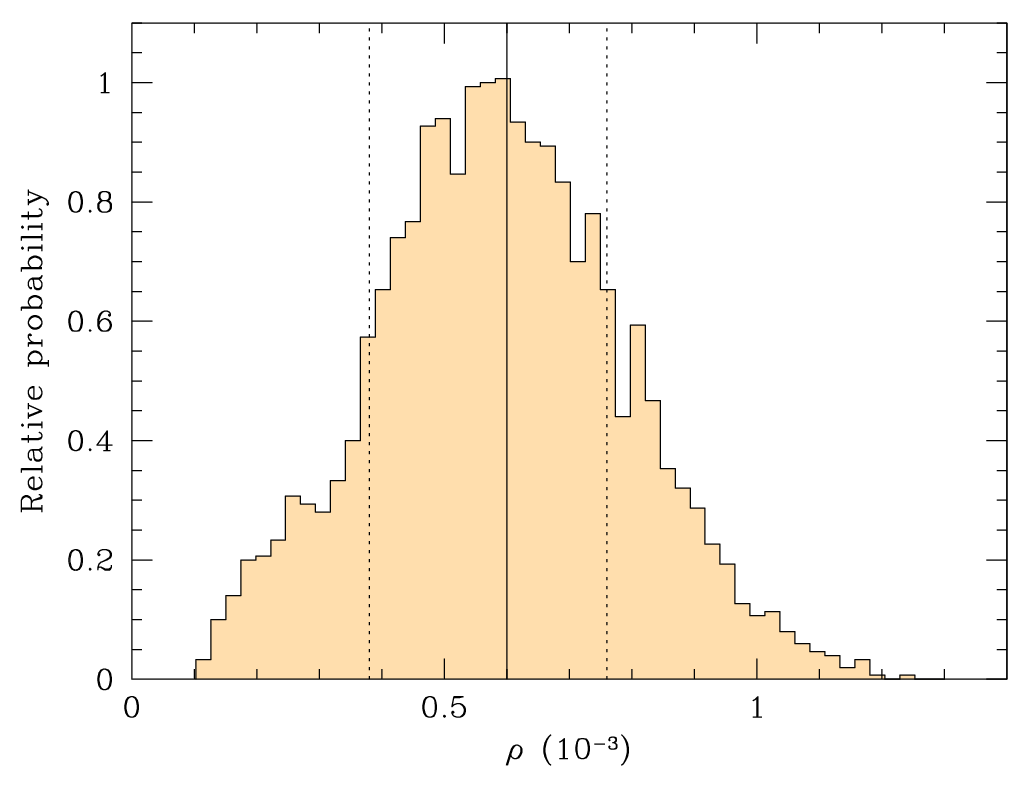}
\caption{
Relative distribution of the normalized source radius.  The solid vertical line indicates 
the median value, and the two dotted lines represent the $1\sigma$ ranges of the distribution.
}
\label{fig:six}
\end{figure}
% --------------------------------------------------------------

\subsection{Triple-lens analysis}\label{sec:three-two}

\citet{Bozza1999} and \citet{Han2001} pointed out that anomalies produced by a triple-lens 
system, composed of three masses $(M_1, M_2, M_3)$, can often be approximated by combining the
anomalies induced by the two binary pairs $M_1$--$M_2$ and $M_1$--$M_3$ through superposition. 
Then, the fact that the two anomaly features in the peak region of the OGLE-2023-BLG-0836 light 
curve are well approximated by two 2L1S models implies the possibility of the lens system 
being a triple system. Consequently, we proceeded with a modeling approach based on a 3L1S
configuration for the lens system.

% Figure 7 ------------------------------------------------------
\begin{figure*}[t]
\centering
\includegraphics[width=12.5cm]{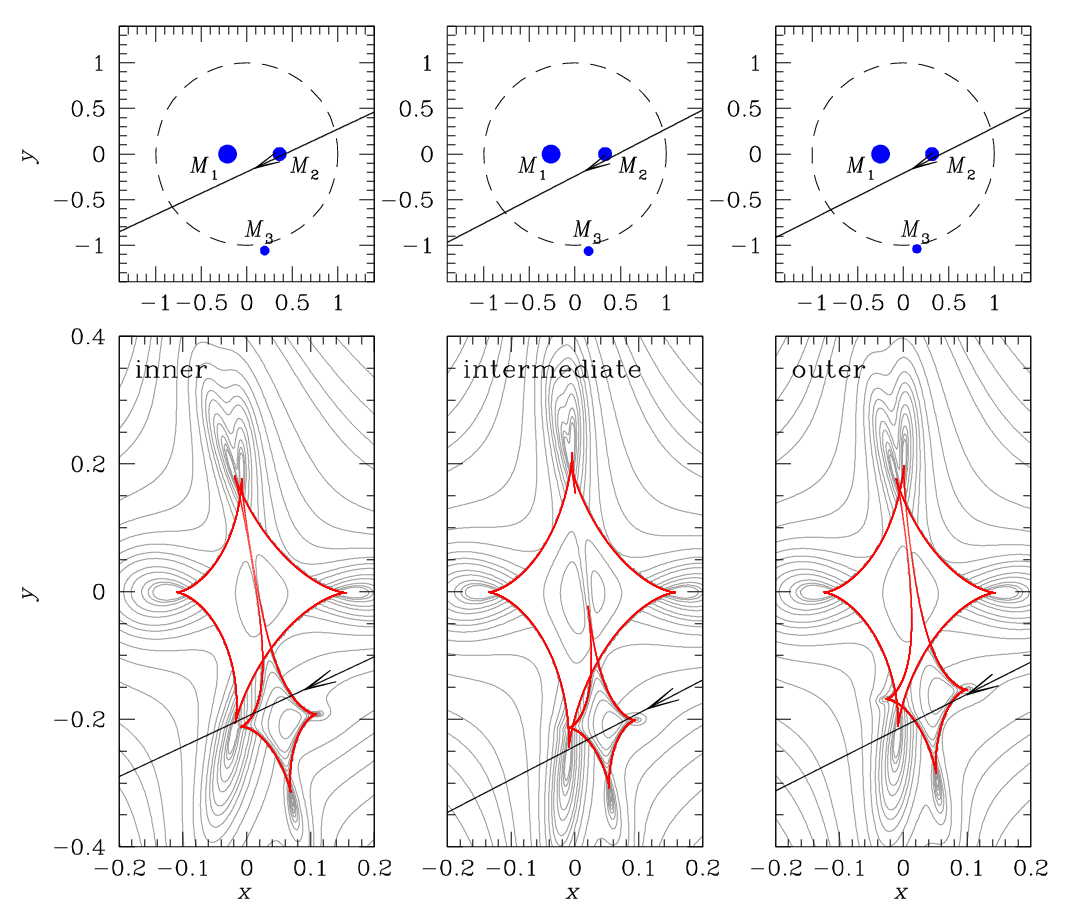}
\caption{
Lens-system configurations of the three degenerate 3L1S solutions. For each solution, the 
lower panel shows the source trajectory, marked by an arrowed line, with respect to the 
caustic, and the upper panel shows the trajectory with respect to the positions of the lens 
components, marked by blue dots with labels $M_1$, $M_2$, and $M_3$. The gray curves 
encompassing the caustic in the lower panels represent equi-magnification contours.  The 
dashed circles in the upper panels represent the Einstein ring. 
}
\label{fig:seven}
\end{figure*}
% --------------------------------------------------------------

The 3L1S configuration corresponds to the case in which an extra lens component, $M_3$, is 
present in addition to the 2L1S configuration.  Incorporating this supplementary lens component
necessitates the addition of extra lensing parameters in the modeling procedure. These parameters 
consist of $(s_3, q_3, \psi)$, which respectively stand for the projected separation and mass 
ratio between $M_1$ and $M_3$, and the orientation angle of $M_3$ as measured from $M_1$--$M_2$ 
axis. In order to distinguish these parameters describing $M_3$ from those describing $M_2$, we 
use to the notations $(s_2, q_2)$ to designate the separation and mass ratio between the 
$M_1$--$M_2$ pair.

The 3L1S modeling was carried out using the following procedure.  In the first step, we
searched for the tertiary lens parameters $(s_3, q_3, \psi)$ via a grid approach using the
parameters of the 2L1S model as initial values for the other parameters. In our analysis, we
used the lensing parameters of the 2L1S solution depicting the cusp-approaching feature.  In 
the second step, we identified local solutions appearing in the $s_3$--$q_3$--$\psi$ parameter 
space, and then refined the individual solutions by gradually minimizing $\chi^2$ of the fit 
using a downhill approach.

Through the 3L1S modeling, we identified three distinct solutions resulting from the ambiguity
in $s_3$. Figure~\ref{fig:five} shows the locations of the individual local solutions in the 
$\Delta\chi^2$ map on the $(\log s_3, \log q_3)$ parameter plane obtained from the grid searches 
for the tertiary lens parameters.  The parameters describing the $M_1$--$M_2$ pair, which lies 
in the ranges of $0.5\lesssim s_2 \lesssim 0.6$ and $0.59\lesssim q_2 \lesssim 0.88$, are similar 
to those of the 2L1S model describing the cusp-approaching feature. Similarly, the parameters 
describing the $M_1$--$M_3$ pair, which lie in the ranges of $1.11\lesssim s_3\lesssim 1.16$ 
and $4.9\times 10^{-3}\lesssim q_3 \lesssim 5.9\times 10^{-3}$, are similar to those of the 
2L1S model describing the caustic-crossing feature. We label the individual local solutions 
as "inner", "intermediate", and "outer" for the reason discussed below.  In the $\Delta\chi^2$ 
map, it appears to be that there exist two local minima with $\log q$ values approximately 
around -2.5. However, during the refinement of the solutions, we found that these two minima 
converge into a single solution.

In Table~\ref{table:two}, we list the full lensing parameters of the three 3L1S solutions 
together with their $\chi^2$ values of the fits and d.o.f. Our analysis reveals a preference 
for the outer solution, with $\chi^2$ differences of 34.0 and 4.2 over the inner and intermediate 
solutions, respectively.  We present the model curve of the outer 3L1S solution in the top panel 
of Figure~\ref{fig:four} and the residuals from all three 3L1S solutions in the lower panels. 
It is worth noting that the normalized source radius is measured based on the finite-source 
constraint, although the measured value carries a substantial uncertainty.  The lensing light 
curve is affected by finite-source effects during both the source crossings over the fold 
caustics induced by the planet and approach to the cusp induced by the binary companion. We 
checked the origin of the $\rho$ constraint by conducting two modeling runs, in which the 
first run was done by excluding the two points near the caustic crossings at HJD$^\prime=
115.685$ and  119.559, and the other run was conducted by excluding the two KMTC points near 
the cusp peak at HJD$^\prime=121.513$ and 121.527.  From these runs, we find that the constraint 
on $\rho$ comes mainly from the two data points observed during the caustic crossings. 
In Figure~\ref{fig:six}, we plot the relative probability of the 
normalized source radius estimated from the 3L1S modeling.

The configurations of the lens systems corresponding to the three 3L1S solutions are presented 
in Figure~\ref{fig:seven}. In all instances, the triple-lens caustic seems to result from the 
combination of two separate binary-lens caustics caused by the $M_1$--$M_2$ and $M_1$--$M_3$ 
pairs. The source first traversed the planetary caustic created by the $M_1$--$M_3$ pair, giving 
rise to the caustic-crossing feature.  Subsequently, the source moved past the lower tip of the 
caustic formed by the $M_1$--$M_2$ pair, leading to the emergence of the cusp-approaching anomaly 
feature. While the fundamental structures of the three solutions bear resemblance to one another, 
there exist subtle distinctions between them. According to the inner and outer solutions, the 
source traversed the two folds of the planetary caustic situated in the inner and outer regions 
between the caustic center and the primary lens, respectively. As we discuss below, the degeneracy 
between the inner and outer solutions is caused by the inner--outer degeneracy \citep{Gaudi1997}. 
In the intermediate solution, the source encountered the inner caustic fold upon entering the 
caustic, and the outer caustic fold while departing from it. We assigned labels to the individual 
solutions based on the side of the caustic that the source crossed.

\citet{Hwang2022} and \citet{Gould2022} showed that the planet separations of a pair of 
solutions resulting from an inner--outer degeneracy follow the relation \begin{equation} 
s^\dagger = \sqrt{s_{\rm in}\times s_{\rm out}} = {\sqrt{u_{\rm anom}^2 + 4} \pm u_{\rm anom} 
\over 2}.  \label{eq4} \end{equation} Here $s_{\rm in}$ and $s_{\rm out}$ respectively denote 
the binary separations of the inner and outer solutions, $u_{\rm anom}^2 = \tau_{\rm anom}^2 
+ u_0^2$, $\tau_{\rm anom} = (t_{\rm anom}- t_0)/\te$, $t_{\rm anom}$ is the time of the 
anomaly.  The symbols "$+$" and "$-$" on the right side of the equation apply to the major-image 
and minor-image perturbations, respectively. In the case of OGLE-2023-BLG-0836, the planet-induced 
anomaly, that is, the caustic-crossing feature, was produced by a major-image perturbation, and 
as a result, the sign associated with this event is "$+$". In order to investigate the origin of 
the degeneracy in the separation $s_3$, we check whether $s_3$ values of the inner and outer 
solutions follow the relation in Eq.~(\ref{eq4}). With the lensing parameters $(t_0, u_0^\prime, 
t_{\rm E}^\prime, t_{\rm anom}, s_{\rm in}^\prime, s_{\rm out}^\prime) = (115.37, 0.22, 46.2, 
117.0, 1.435, 1.408)$, we find that the geometric mean $(s_{\rm in}\times s_{\rm out})^{1/2}\sim 
1.124$ well matches the value $[(u_{\rm anom}^2 + 4)^{1/2} + u_{\rm anom}]/2 = 1.119$. This 
confirms that the similarity between the model curves of the inner and outer solutions stems 
from the inner--outer degeneracy. In computing $s^\dagger$, we used the values of the lensing 
parameters normalized to the Einstein radius corresponding to sum of $M_1$ and $M_3$, that is, 
$(u_0^\prime, t_{\rm E}^\prime, s_{\rm in}^\prime, s_{\rm out}^\prime)\equiv (u_0f, \te/f, 
s_{\rm in} f, s_{\rm out} f)$, where $f=(1+q_3)^{1/2}$.

% Figure 8 ------------------------------------------------------
\begin{figure}[t]
\includegraphics[width=\columnwidth]{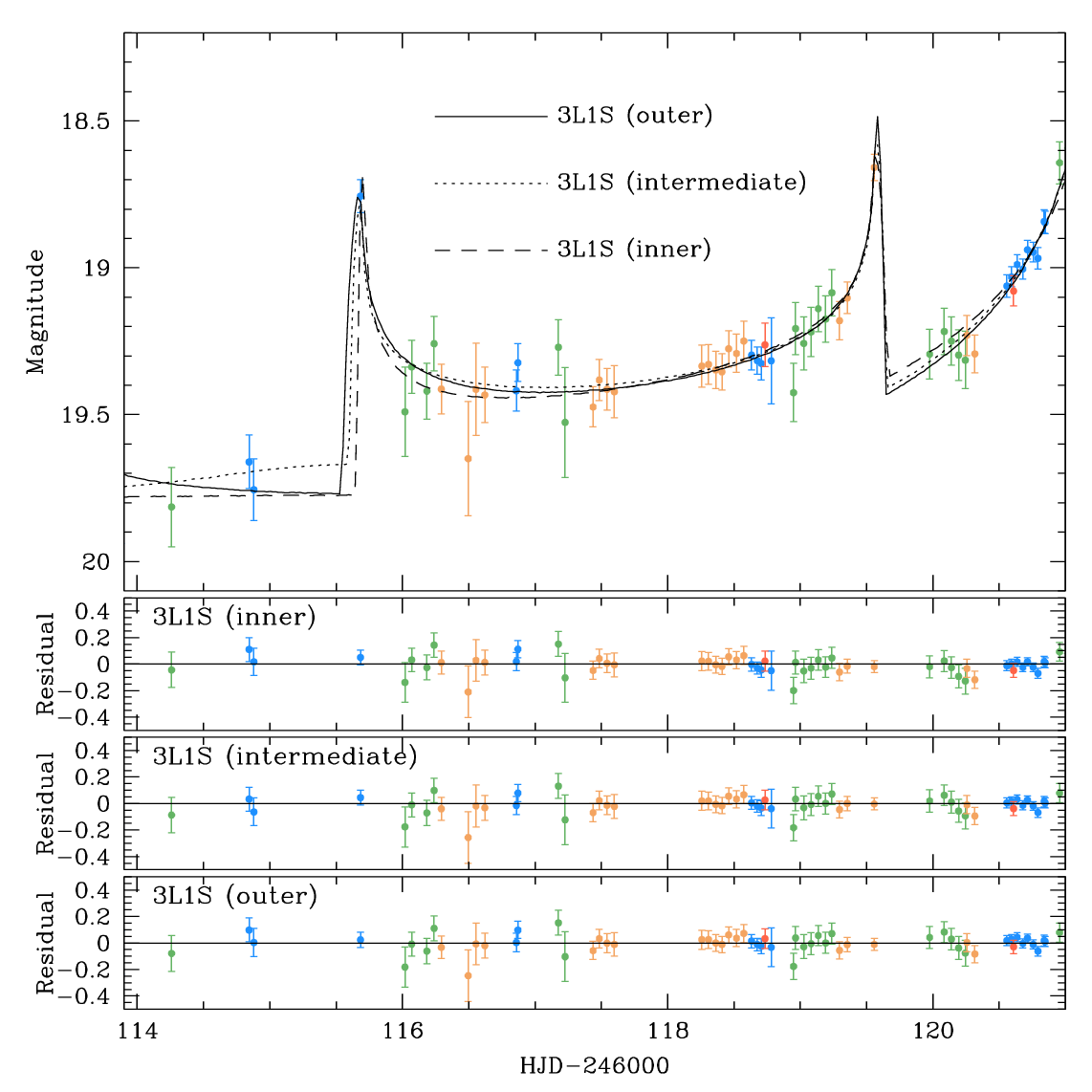}
\caption{
Zoom-in view around the caustic-crossing feature.  Model curves of the three degenerate 
solutions (outer, intermediate, and inner) are drawn in the top panel, and the residuals 
from the models are shown in the lower panels.
}
\label{fig:eight}
\end{figure}
% --------------------------------------------------------------

After identifying the origin of the  degeneracy between the inner and outer solutions,
we further investigate the origin of the degeneracy involving the intermediate solution 
and the other solutions. It is worth noting that \citet{Shin2023} and \citet{Han2023c} 
independently reported the presence of such degeneracies in their respective analyses of 
the planetary lensing events KMT-2016-BLG-1751 and KMT-2023-BLG-0469.  Upon closer 
examination of the caustic-crossing feature, we found that the degeneracy between the 
intermediate solution and the other solutions is an accidental one stemming from inadequate 
caustic coverage.  This is depicted in Figure~\ref{fig:eight}, which provides a detailed 
view around the caustic-crossing feature.  The figure illustrates that the source magnitudes 
just before the caustic entrance and just after the caustic exit for the inner and outer 
solutions are remarkably similar, suggesting an inherent degeneracy.  On the contrary, 
the source magnitude just before the caustic entrance for the intermediate solution is 
approximately 0.1 magnitude brighter than what is expected from the other solutions, 
indicating that this degeneracy is a chance occurrence.  If the light curve in this 
specific region had been more thoroughly covered, the degeneracy between the intermediate 
solution and the other solutions could have been lifted. The issue of degeneracies arising 
due to insufficient coverage of certain parts of caustic-crossing features was explored by 
\citet{Skowron2018}.

\subsection{Binary-lens binary-source analysis}\label{sec:three-three}

As illustrated in the lensing event KMT-2021-BLG-0240 \citep{Han2022c}, anomalies arising 
from a 3L1S system can, on occasion, be confounded with anomalies resulting from 2L2S systems.
Hence, we conducted a more thorough examination to ascertain whether the observed anomalies 
in the lensing light curve of OGLE-2023-BLG-0836 could be explained by a 2L2S interpretation.

% Figure 9 ------------------------------------------------------
\begin{figure}[t]
\includegraphics[width=\columnwidth]{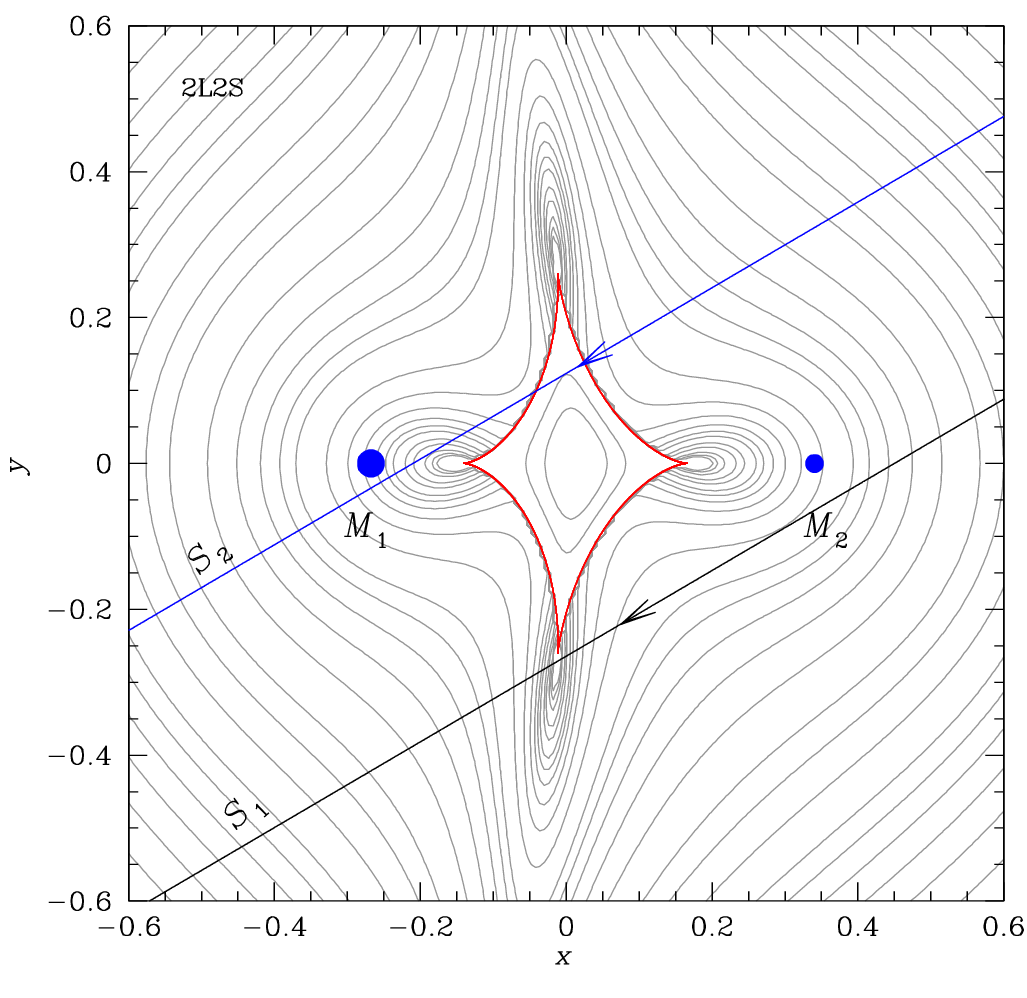}
\caption{
Lens-system configuration of the 2L2S solution. Notations are same as those in Fig.~\ref{fig:six} 
except that there are two source trajectories. The black and blue lines represent the trajectories 
of the primary and secondary source stars, respectively. 
}
\label{fig:nine}
\end{figure}
% --------------------------------------------------------------

The 2L2S configuration involves the presence of an additional source alongside the 2L1S 
configuration. We designate the primary and secondary source stars as $S_1$ and $S_2$, 
respectively. To account for the additional source, it is necessary to incorporate extra 
lensing parameters into the modeling process. These additional parameters encompass $(t_{0,2}, 
u_{0,2}, \rho_2, q_F)$, which respectively represent the time and separation of the closest 
approach of $S_2$ to the lens, the normalized source radius of $S_2$, and the flux ratio 
between $S_2$ and $S_1$. Concurrently, we employ the notations $(t_{0,1}, u_{0,1}, \rho_1)$ 
to define the parameters describing the approach of $S_1$ to the lens. During the modeling 
process, we seek the 2L2S parameters by exploring various trajectories for $S_2$, building 
upon the 2L1S solution that describes the caustic-crossing feature.

In Table~\ref{table:three}, we list the best-fit lensing parameters of the 2L2S solution. The 
lens-system configuration corresponding to the solution is shown in Figure~\ref{fig:nine}. 
In this illustration, the paths of the primary and secondary source stars are indicated by 
black and blue lines, respectively, and they are labeled as $S_1$ and $S_2$. The model curve 
and residual of the 2L2S solution in the region of the anomaly are presented in 
Figure~\ref{fig:four}, showing that the model approximately describes the anomaly features.

% Figure 10 ------------------------------------------------------
\begin{figure}[t]
\includegraphics[width=\columnwidth]{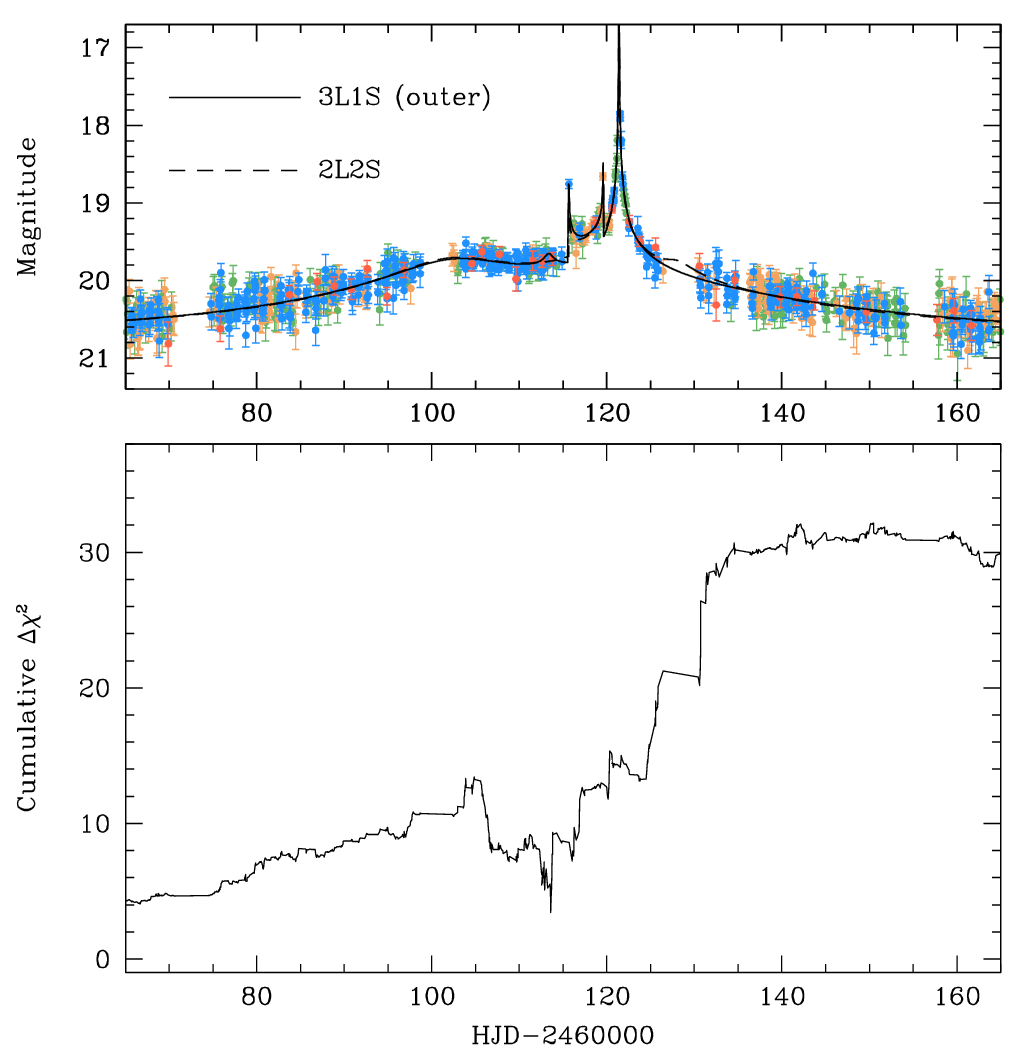}
\caption{
Cumulative distribution of $\chi^2$ difference between the 2L2S and 3L1S solutions. The light 
curve in the upper panel is provided to illustrate the region of disparity in the fit. 
}
\label{fig:ten}
\end{figure}
% --------------------------------------------------------------

% Table 3 ------------------------------------------------
\begin{table}[t]
%\footnotesize
%\small
%\centering
\caption{Best-fit parameters of 2L2S solution.\label{table:three}}
%\begin{tabular}{lllllll}
\begin{tabular*}{\columnwidth}{@{\extracolsep{\fill}}lllcc}
\hline\hline
\multicolumn{1}{c}{Parameter}     &
\multicolumn{1}{c}{Value}    \\
\hline
 $\chi^2$/d.o.f.          &  $1618.1/1619        $   \\  %$1620.8/1619        $
 $t_{0,1}$ (HJD$^\prime$) &  $  113.80 \pm 0.22  $   \\  %$60115.37 \pm 0.17  $
 $u_{0,1}$                &  $0.227  \pm 0.012   $   \\  %$0.1773 \pm 0.0073  $
 $t_{0,2}$ (HJD$^\prime$) &  $120.34 \pm 0.12    $   \\  %$58.98 \pm 1.76     $
 $u_{0,2}$                &  $-0.1064 \pm 0.0043 $   \\  %$0.5717 \pm 0.0082  $
 $\te$ (days)             &  $51.12 \pm 2.32     $   \\  %$0.593 \pm 0.053    $
 $s$                      &  $0.608  \pm 0.012   $   \\  %$-0.439 \pm 0.013   $
 $q$                      &  $0.786  \pm 0.036   $   \\  %$1.1344 \pm 0.0073  $
 $\alpha$ (rad)           &  $-0.531 \pm 0.017   $   \\  %$4.88 \pm 0.82      $
 $\rho_1$ ($10^{-3}$)     &  $0.74 \pm 0.46      $   \\  %$5.081 \pm 0.011    $
 $\rho_2$ ($10^{-3}$)     &  $0.47 \pm 0.21      $   \\  %$0.64 \pm 0.19      $
 $q_F$                    &  $0.0797 \pm 0.0076  $   \\  %$0.64 \pm 0.19      $
\hline
\end{tabular*}
%\tablefoot{ ${\rm HJD}^\prime = {\rm HJD}- 2460000$.  }
\end{table}
% --------------------------------------------------------

Although the 2L2S model offers an approximate description of the anomaly features, it is 
found that the model exhibits a relatively less accurate fit to the data when compared 
to the 3L1S model. This can be seen in Figure~\ref{fig:ten}, in which the cumulative 
distribution of $\chi^2$ difference between the 2L2S and 3L1S solutions, that is, 
$\Delta\chi^2 = \chi^2_{\rm 2L2S}-\chi^2_{\rm 3L1S}$, is presented. From the distribution, 
it is found that the 2L2S solution yields a poorer fit in the region after the 
cusp-approaching feature.  Overall, it is found that the 2L2S solution is less preferred 
by $\Delta\chi^2  = 31.3$ when compared to the 3L1S solution. As a result, we dismiss the 
2L2S interpretation for the anomaly.

\section{Source star and angular Einstein radius}\label{sec:four}

In this section, we specify the source star of the lensing event and estimate the angular Einstein
radius. The source star is specified by estimating its color and magnitude after being corrected 
for reddening and extinction. With the angular source radius $\theta_*$ deduced from the color and
magnitude together with the normalized source radius $\rho$ measured from the modeling, the angular
Einstein radius is estimated as
\begin{equation}
\thetae = {\theta_* \over \rho}.
\label{eq5}
\end{equation}

\begin{figure}[t]
\includegraphics[width=\columnwidth]{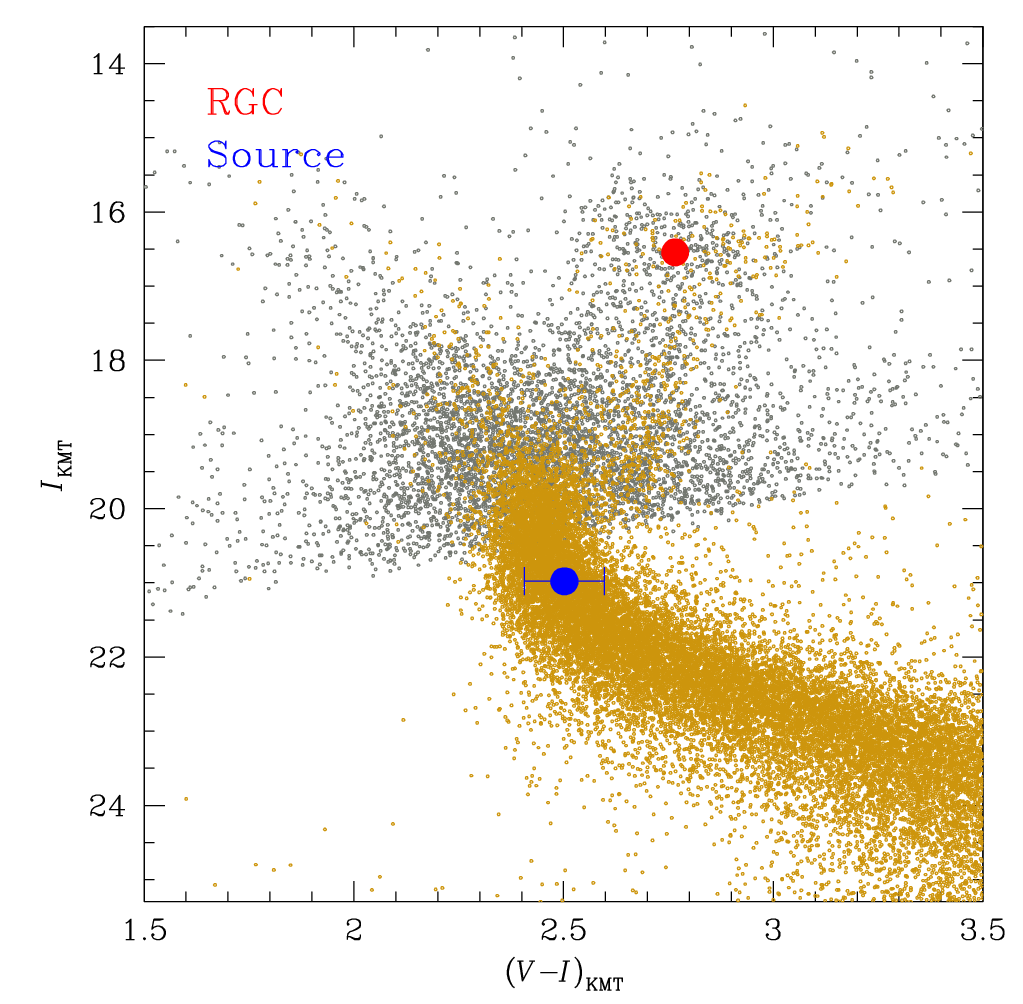}
\caption{
Positions of the source and red giant clump (RGC) centroid in the instrumental color-magnitude 
diagram (CMD). The CMD is created by merging observations from KMTC (gray dots) and HST (brown 
dots).
}
\label{fig:eleven}
\end{figure}

Figure~\ref{fig:eleven} shows the location of the source in the instrumental color-magnitude diagram 
(CMD) of stars lying near the source. The CMD was created by merging two sets of CMDs: one set 
generated from pyDIA photometry \citep{Albrow2017} of stars in the KMTC image and the other set 
pertains to stars in Baade's window observed using the Hubble Space Telescope \citep{Holtzman1998}. 
The alignment of these two CMDs was achieved by using the centroids of the red giant clump (RGC) 
in the individual CMDs. We used the combined CMD because the $V$-band source magnitude could not 
be determined although the $I$-band magnitude of the source was measured. As a result, the color 
of the source was estimated as the median value observed in the main-sequence branch of the 
combined CMD corresponding to the measured $I$-band magnitude.

% Figure 11 ------------------------------------------------------
% --------------------------------------------------------------

To estimate the de-reddened source color and magnitude, denoted as $(V - I, I)_0$, from their
corresponding instrumental values, denoted as $(V - I, I)$, we employed the \citet{Yoo2004} 
method. This method utilizes the RGC centroid, for which its de-reddened color and magnitude 
$(V - I, I)_{{\rm RGC},0} = (1.060, 14.308)$ have been established by previous studies 
\citep{Bensby2013, Nataf2013}, as a reference point for calibration, that is, 
\begin{equation}
(V-I, I)_{s,0} = (V-I, I)_{{\rm RGC},0} + \Delta(V-I, I). 
\label{eq6}
\end{equation}
Here $(V-I, I)_{\rm RGC}$ represent the instrumental color and magnitude of the RGC centroid, 
and $\Delta(V-I, I) = (V-I, I) - (V-I, I)_{\rm RGC}$ indicate the offsets in color and magnitude 
between the source and RGC centroid. In Table~\ref{table:four}, we list the measured values 
$(V-I, I)$, $(V-I, I)_{\rm RGC}$, $(V-I, I)_{{\rm RGC},0}$, and $(V-I, I)_0$.  From the estimated 
de-reddened color and magnitude, it is found that the source is a late G-type main-sequence star.

For the estimation of the angular source radius, we first converted the measured $V - I$ color 
into $V - K$ color using the color-color relation established by \citet{Bessell1988}, and 
subsequently calculated the angular source radius by applying the relation between $V - K$ and 
$\theta_*$ provided by \citet{Kervella2004}. The derived value of the angular source radius is
\begin{equation}
\theta_* = (0.618 \pm 0.073)~\mu{\rm as}.
\label{eq7}
\end{equation}
This yields the angular Einstein radius and relative lens-source proper motion of
\begin{equation}
\thetae = (0.97 \pm 0.31)~{\rm mas}. 
\label{eq8}
\end{equation}
and
\begin{equation}
\mu = {\thetae \over \te} = (5.80 \pm 1.86)~{\rm mas}~{\rm yr}^{-1}
\label{eq9}
\end{equation}
respectively. We note that the uncertainties associated with $\thetae$ and $\mu$ are relatively 
large, primarily stemming from the large uncertainty in $\rho$.

A degeneracy between two competing interpretations with widely different lensing parameters 
of $(\te, \rho)$ can occasionally be lifted from the resulting values of the relative lens-source 
proper motion if one model results in $\mu$ value that is relatively disfavored by the Galactic 
model. We checked the feasibility of this method by additionally estimating the relative proper 
motion expected from the 2L2S model. The estimated value of $\mu_{\rm 2L2S}\sim 6.0$~mas/yr, which 
is not only similar to the value for the 3L1S model but also very typical value for a Galactic 
lensing event. Therefore, the constraint from the estimated relative proper motion cannot used 
for distinguishing the two interpretations.

% Table 4 ------------------------------------------------
\begin{table}[t]
%\small
%\centering
\caption{Source parameters.\label{table:four}}
%\begin{tabular}{lllllll}
\begin{tabular*}{\columnwidth}{@{\extracolsep{\fill}}lllcc}
\hline\hline
\multicolumn{1}{c}{Parameter}    &
\multicolumn{1}{c}{Value}        \\
\hline
 $(V-I, I)$                 &  $(2.502 \pm 0.095, 20.978 \pm 0.010)$     \\
 $(V-I, I)_{\rm RGC}$       &  $(2.766, 16.543)                    $     \\
 $(V-I, I)_{{\rm RGC},0}$   &  $(1.060, 14.308)                    $     \\
 $(V-I, I)_0$               &  $(0.796 \pm 0.095, 18.743 \pm 0.010)$     \\
\hline
\end{tabular*}
%\tablefoot{ ${\rm HJD}^\prime = {\rm HJD}- 2450000$.  }
\end{table}
% --------------------------------------------------------

\section{Physical lens parameters}\label{sec:five}

In this section, we estimate the physical parameters of the lens system. The lens parameters 
of the mass $M$ and distance $\dl$ are constrained by three lensing observables: the event 
time scale $\te$, the angular Einstein radius $\thetae$, and the microlens parallax $\pie$. 
The microlens parallax is defined as the ratio of the relative lens-source parallax 
$\pi_{\rm rel}$ to the angular Einstein
radius, that is, 
\begin{equation}
\pie = {\pi_{\rm rel} \over \thetae};\qquad
\pi_{\rm rel} = {\rm au} \left( {1\over \dl}-{1\over D_{\rm S}} \right),
\label{eq10}
\end{equation}
where $\ds$ denotes the distance to the source. With the measurements of all these observables, 
the mass and distance to the lens are uniquely determined as 
\begin{equation}
M = {\thetae \over \kappa\pie};\qquad
\dl = {{\rm au} \over \pie\thetae + \pi_{\rm S}},
\label{eq11}
\end{equation}
where $\kappa =4G/(c^2{\rm au})$ and $\pi_{\rm S}={\rm au}/\ds$ denotes the parallax of the 
source \citep{Gould2000}.  For OGLE-2023-BLG-0836, while the observables $\te$ and $\thetae$ 
were constrained, the accurate determination of microlens parallax was hindered by limited 
photometric data precision. Consequently, we estimate $M$ and $\dl$ by conducting a Bayesian 
analysis, leveraging the constraints provided by the measured observables $\te$ and $\thetae$ 
together with the priors of the physical and dynamical distributions and mass function of lens 
objects in the Galaxy.

% Figure 12 ------------------------------------------------------
\begin{figure}[t]
\includegraphics[width=\columnwidth]{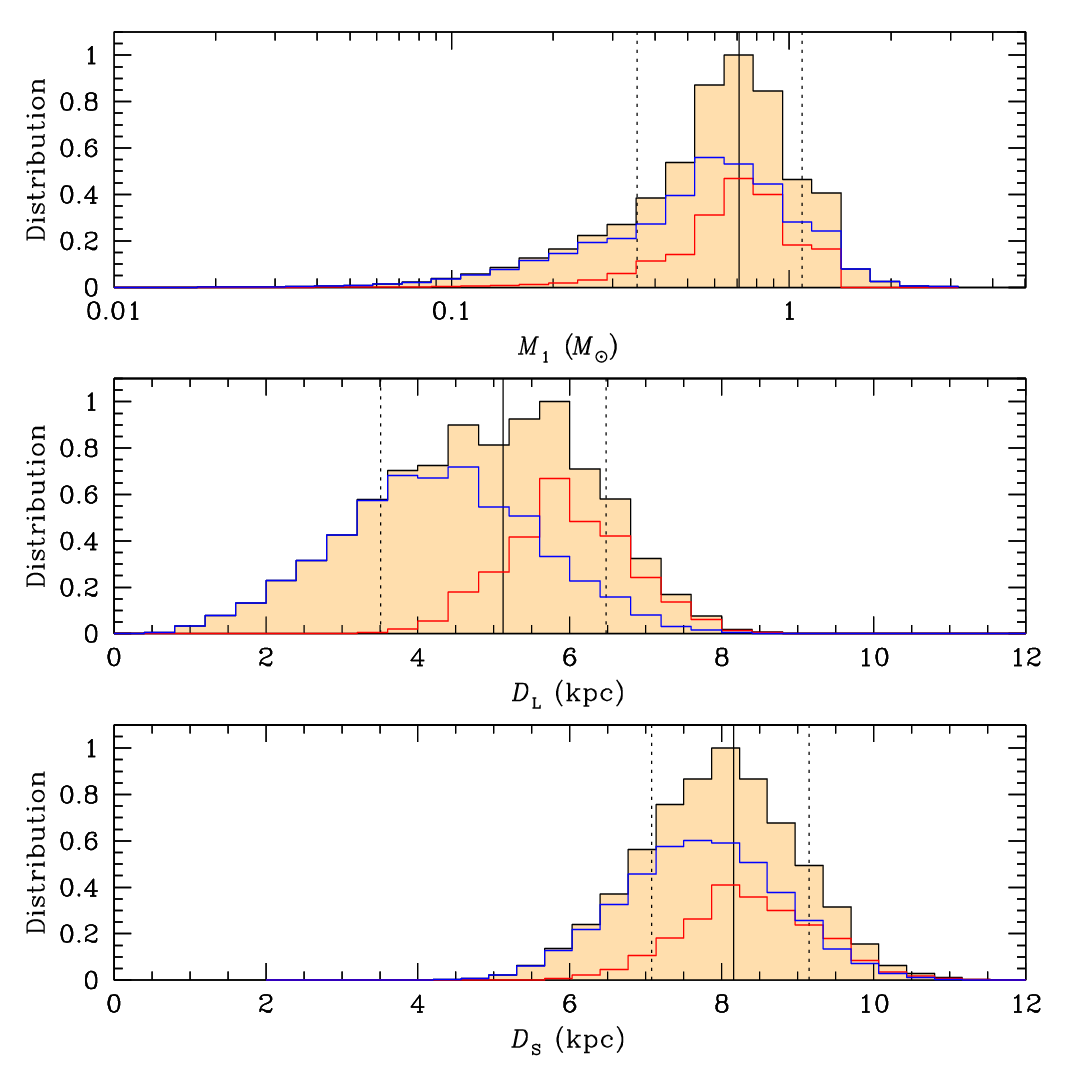}
\caption{
Bayesian posteriors of the primary lens mass ($M_1$) and the distances to the lens
($\dl$) and source ($\ds$) of the lensing event OGLE-2023-BLG-0836. Within each panel,
the event contributions from the disk and bulge lenses are illustrated by blue and red
curves, and the black curve represents the combined distribution of the two lens
populations. The solid vertical line marks the median value, and the dotted lines indicate
the $1\sigma$ range of the posterior distribution. 
}
\label{fig:twelve}
\end{figure}
% --------------------------------------------------------------

The Bayesian analysis was initiated by generating a large number of synthetic events through 
a Monte Carlo simulation. In the simulation, the physical parameters of the lens mass were 
derived from a model mass function, and the distances to the lens and source as well as the 
relative proper motion between them were obtained from a Galaxy model. We adopted the mass 
function model proposed by \citet{Jung2018} and the Galaxy model introduced by \citet{Jung2021}.  
Using physical parameters $(M_i, D_{{\rm L},i}, D_{{\rm S},i}, \mu_i)$ for each synthetic lensing 
event, we computed the corresponding lensing observables, specifically the event time scale and 
angular Einstein radius, according to the following relations: 
\begin{equation}
t_{{\rm E},i} = {\theta_{{\rm E},i} \over \mu_i};\qquad
\theta_{{\rm E},i} = (\kappa M_i \pi_{{\rm rel},i})^{1/2}.
\label{eq12}
\end{equation}
Then, the posteriors of the lens mass and distance were subsequently constructed by assigning 
a weight to each event of 
\begin{equation}
w_i = \exp \left(-{\chi_i^2 \over 2}\right);\qquad
\chi_i^2 = 
{ (t_{{\rm E},i} - \te)^2\over \sigma^2(\te)} + 
{ (\theta_{{\rm E},i} - \thetae)^2\over \sigma^2(\thetae)}.
\label{eq13}
\end{equation}
Here, $(\te, \thetae)$ represent the measured values of the observables, and $[\sigma(\te), 
\sigma(\thetae)]$  denote their respective uncertainties.

In Figure~\ref{fig:twelve}, we present the Bayesian posteriors of the primary lens mass $M_1$ 
and the distances to the lens and source, $\dl$ and $\ds$, of OGLE-2023-BLG-0836. In 
Table~\ref{table:five}, we list the masses of the individual lens components $(M_1, M_2, 
M_3)$, distance to the lens, and the projected separations of the individual lens companions 
from the primary, $(a_{\perp,2}, a_{\perp,3})$. The physical parameters were derived based on 
the observables of the outer solution, which provided the best fit to the data. Given the 
similarity of observables, the physical parameters derived from the other solutions are similar 
to the presented values. For each physical parameter, we provide the median of the Bayesian 
posterior as a representative value and estimate the uncertainty range as the 16\% and 84\% of 
the distribution. The least massive component of the lens has a mass $M_3 \sim 4.4~M_{\rm J}$, 
classifying it as a giant planet. The masses of the other lens components $M_1 \sim 0.71~M_\odot$ 
and $M_2 \sim 0.56~M_\odot$ correspond to those of mid and late K-type main-sequence stars. Given 
that the planetary separation between $M_1$ and $M_3$, $a_{\perp,3} \sim 3.7$~au, is greater than 
that between the $M_1$ and $M_2$ binary pair, $a_{\perp,2} \sim 1.9$~au, it is likely that the 
planet orbits both binary components. However, the possibility of the planet orbiting one of the 
binary components cannot be entirely ruled out because of the projected nature of the separations. 
The estimated distance to the lens is $\dl \sim 5.1$~kpc, and the probabilities for the lens lying 
in the disk and bulge are 66\% and 34\%, respectively.

% Table 5 ------------------------------------------------
\begin{table}[t]
%\small
%\centering
\caption{Physical lens parameters.\label{table:five}}
%\begin{tabular}{lllllll}
\begin{tabular*}{\columnwidth}{@{\extracolsep{\fill}}lllcc}
\hline\hline
\multicolumn{1}{c}{Parameter}    &
\multicolumn{1}{c}{Value}        \\
\hline
$M_1$ ($M_\odot$)      &   $0.71^{+0.38}_{-0.36}$    \\    [0.8ex]
$M_2$ ($M_\odot$)      &   $0.56^{+0.30}_{-0.28}$    \\    [0.8ex]
$M_3$ ($M_{\rm J}$)    &   $4.36^{+2.35}_{-2.18}$    \\    [0.8ex]
$\dl$ (kpc)            &   $5.12^{+1.36}_{-1.61}$    \\    [0.8ex]
$a_{\perp,2}$ (au)     &   $1.88^{+0.50}_{-0.59}$    \\    [0.8ex]
$a_{\perp,3}$ (au)     &   $3.70^{+0.98}_{-1.16}$    \\    [0.8ex]

\hline
\end{tabular*}
%\tablefoot{ ${\rm HJD}^\prime = {\rm HJD}- 2450000$.  }
\end{table}
% --------------------------------------------------------

\section{Summary and conclusion}\label{sec:six}

We have conducted an analysis of the peculiar lensing event OGLE-2023-BLG-0836, for which the
light curve is characterized by two distinctive anomaly features produced by a caustic crossing and a cusp
approach of a source. Despite the comprehensive exploration of the parameter space, we were
unable to identify a binary-lens solution capable of sufficiently explaining both features within the
anomaly.

From the analysis with sophisticated model prompted by the fact that each anomaly feature can be
approximated by a 2L1S model, we have arrived at the conclusion that a triple-mass lens system is
imperative to account for the observed anomaly pattern in the lensing light curve. A binary-lens
binary-source interpretation could also offer an approximate explanation for the anomaly pattern,
but this interpretation was rejected with a high degree of statistical confidence. Through the
detailed triple-lens modeling, we identified three distinct solutions resulting from the degeneracy in
the separation between the primary and the least massive companion of the lens.

Through a Bayesian analysis using the measured observables of the event time scale and angular
Einstein radius, we determined that the least massive component of the lens has a planetary mass
of $4.36^{+2.35}_{-2.18}~M_{\rm J}$. This planet orbits within a stellar binary system composed of two
stars with masses $0.71^{+0.38}_{-0.36}~M_\odot$ and $0.56^{+0.30}_{-0.28}~M_\odot$. This lensing event
signifies the sixth occurrence of a microlensing planetary system in which a planet belongs to a
stellar binary system.

\begin{acknowledgements}
Work by C.H. was supported by the grants of National Research Foundation of Korea (2019R1A2C2085965). 
% Yee
J.C.Y. and I.-G.S. acknowledge support from U.S. NSF Grant No. AST-2108414. 
% Yossi Shvartzvald 
Y.S. acknowledges support from BSF Grant No. 2020740.
% KMTNet
This research has made use of the KMTNet system operated by the Korea Astronomy and Space
Science Institute (KASI) at three host sites of CTIO in Chile, SAAO in South Africa, and SSO in
Australia. Data transfer from the host site to KASI was supported by the Korea Research
Environment Open NETwork (KREONET). This research was supported by KASI under the R\&D
program (project No. 2023-1-832-03) supervised by the Ministry of Science and ICT.
% Chinese researcher 
W.Zang, H.Y., S.M., R.K., J.Z., and W.Zhu acknowledge support by the National Natural Science
Foundation of China (Grant No. 12133005).
W.Zang acknowledges the support from the Harvard-Smithsonian Center for Astrophysics through
the CfA Fellowship. 
% OGLE 
The OGLE has received funding from the National Science Centre, Poland, grant MAESTRO
2014/14/A/ST9/00121 to AU.
\end{acknowledgements}


\begin{thebibliography}{}
% -------------
%\bibitem[Albrow et al.(2000)]{Albrow2000} Albrow, M. D., Beaulieu, J.-P., Caldwell, J. A. R., et al. 2000, \apj, 534, 894
%\bibitem[Albrow et al.(2009)]{Albrow2009} Albrow, M., Horne, K., Bramich, D.~M., et al.\ 2009, \mnras, 397, 2099
\bibitem[Albrow(2017)]{Albrow2017} Albrow, M.\ 2017, MichaelDAlbrow/pyDIA: Initial Release on Github,Versionv1.0.0, Zenodo, doi:10.5281/zenodo.268049
%\bibitem[An(2005)]{An2005} An, J. H. 2005, \mnras, 356, 1409
\bibitem[Bennett et al.(2016)]{Bennett016}   Bennett, D. P., Rhie, S. H., Udalski, A. et al. 2016, \aj, 152, 125
\bibitem[Bensby et al.(2013)]{Bensby2013} Bensby, T., Yee, J.~C., Feltzing, S., et al.\ 2013, \aap, 549, A147
%\bibitem[Bond et al.(2001)]{Bond2001}  Bond, I. A., Abe, F.;,Dodd, R. J., et al. 2001, \mnras, 327, 868
\bibitem[Bessell \& Brett(1988)]{Bessell1988} Bessell, M.~S., \& Brett, J.~M. 1988, \pasp, 100, 1134
\bibitem[Bozza(1999)]{Bozza1999} Bozza, V. 1999, \aap, 348, 311 
%\bibitem[Dominik(1999)]{Dominik1999} Dominik, M. 1999, \aap, 349, 108
\bibitem[Doran \& Mueller(2004)]{Doran2004}    Doran, M., \& Mueller, C. M. 2004, J. Cosmology Astropart. Phys., 09, 003
\bibitem[Gaudi \& Gould(1997)]{Gaudi1997}   Gaudi, B. S., \& Gould, A. 1997, \apj, 486, 85
%\bibitem[Gaudi et al.(2008)]{Gaudi2008}   Gaudi, B. S., Bennett, D. P., Udalski, A., et al. 2008, Science, 319, 927
%\bibitem[Gould(1992)]{Gould1992}  Gould, A. 1992, \apj, 392, 442
\bibitem[Gould(2000)]{Gould2000}    Gould, A. 2000, \apj, 542, 785
\bibitem[Gould et al.(2014)]{Gould2014}   Gould, A., Udalski, A., Shin, I.-G., et al. 2014, Science, 345, 46
\bibitem[Gould et al.(2022)]{Gould2022}   Gould, A., Han, C., Weicheng, Z., et al. 2022, \aap, 664, A13 
%\bibitem[Gould et al.(2023)]{Gould2023}    Gould, A., Ryu, Y.-H., Yee, J. C., et al. 2023, \aj, 166, 100 
\bibitem[Griest \& Hu(1992)]{Griest1992}  Griest, K., \& Hu, W. 1992, \apj, 397, 362  
%\bibitem[Griest \& Safizadeh(1998)]{Griest1998}  Griest, K., \& Safizadeh, N. 1998, \apj, 500, 37
\bibitem[Han(2001)]{Han2001}  Han, C., Chang, H.-Y., An, J. H., \& Chang, K. 2001, \mnras, 328, 986  
%\bibitem[Han \& Gould(2003)]{Han2003}  Han, C., \& Gould, A. 2003, \apj, 592, 172
%\bibitem[Han \& Gaudi(2008)]{Han2008}  Han, C., \& Gaudi, B. S. 2008, \apj, 689, 53
\bibitem[Han et al.(2017)]{Han2017}    Han, C., Udalski, A., Gould, A., et al. 2017, \aj, 154, 223
\bibitem[Han et al.(2019)]{Han2019}    Han, C., Bennett, D. P., Udalski, A., et al. 2019, \aj, 158, 114      %(Han2019)  %OB-18-BLG-1011Lb,c:  
\bibitem[Han et al.(2020a)]{Han2020a}  Han, C., Lee, C.-U., Udalski, A., et al. 2020a, \aj, 159, 48        %(Han2020a) %OGLE-2018-BLG-1700L:  
\bibitem[Han et al.(2020b)]{Han2020b}  Han, C., Kim, D., Jung Y. K., et al. 2020b, \aj, 160, 17            %(Han2020b) %KMT-2019-BLG-1953�  
\bibitem[Han et al.(2020c)]{Han2020c}  Han, C., Udalski, A., Lee, C.-U. et al. 2020c, \aj, 162, 203
\bibitem[Han et al.(2021a)]{Han2021a}  Han, C., Lee, C.-U., Ryu, Y.-H., et al. 2021a,\ \aap, 649, A91      %(Han2021a) %KMT-2019-BLG-0797:  
\bibitem[Han et al.(2021b)]{Han2021b}  Han, C., Udalski, A., Kim, D., et al. 2021b, \aj, 161, 270          %(Han2021b) %KMT-2019-BLG-1715:  
\bibitem[Han et al.(2021c)]{Han2021c}  Han, C., Albrow, M. D., Chung S.-J., et al. 2021c, \aap, 652, A145  %(Han2021c) KB-18-1743:  
\bibitem[Han et al.(2021d)]{Han2021d}  Han, C., Udalski, A., Lee, C.-U., et al. 2021c, \aj, 162, 203       %(Han2021d) �OGLE-2019-BLG-0304:  
\bibitem[Han et al.(2022a)]{Han2022a}  Han, C., Gould, A., Bond, I. A., et al. 2022a, \aap, 662, A70       %(Han2022a) �KB-2021-BLG-1077L:  
\bibitem[Han et al.(2022b)]{Han2022b}  Han, C., Gould, A., Kim, D., et al. 2022b, \aap, 663, A145          %(Han2022b) �KB-21-BLG-1898:  
\bibitem[Han et al.(2022c)]{Han2022c}  Han, C., Kim, D., Yang, H., et al. 2022c, \aap, 664, A114           %(Han2022c) �KMT-2021-BLG-0240: 
%\bibitem[Han et al.(2022d)]{Han2022d}  Han, C., Ryu, Y.-H., Shin, I.-Gu, et al. 2022d, \aap, 667, A64      %(Han2022d) BD events 
\bibitem[Han et al.(2023a)]{Han2023a}  Han, C., Udalski, A., Jung Y. K., et al. 2023a, \aap, 670, A172     %(Han2023a) �OGLE-2018-BLG-0584 and KMT-2018-BLG-2119:  
\bibitem[Han et al.(2023b)]{Han2023b}  Han, C., Jung, Y. K., Gould, A., et al. 2023b, \aap, 672, A8        %(Han2023b) �KMT-21-BLG-1122L:  
\bibitem[Han et al.(2023c)]{Han2023c}  Han, C., Jung, Y. K., Bond, I. A., et al. 2023c, \aap, submitted       
%\bibitem[Han et al.(2023d)]{Han2023d}  Han, C., Jung, Y. K., Bond, I. A., et al. 2023d, \aap, 675, A71     %(Han2023d) arXiv:2307.04921 BD events 
\bibitem[Holtzman et al.(1998)]{Holtzman1998}   Holtzman, J. A., Watson, A. M., Baum, W. A., et al. 1998, \aj, 115, 1946 
\bibitem[Hwang et al.(2022)]{Hwang2022}   Hwang, K.-H., Zang, W., Gould, A., et al. 2022, \aj, 163, 43
\bibitem[Jung et al.(2018)]{Jung2018}    Jung, Y. K., Udalski, A., Gould, A., et al. 2018, \aj, 155, 219
\bibitem[Jung et al.(2021)]{Jung2021}    Jung, Y. K., Han, C., Udalski, A., et al. 2021, \aj, 161, 293
\bibitem[Kervella et al.(2004)]{Kervella2004} Kervella, P., Th\'evenin, F., Di Folco, E., \& S\'egransan, D.\ 2004, \aap, 426, 29
\bibitem[Kim et al.(2016)]{Kim2016} Kim, S.-L., Lee, C.-U., Park, B.-G., et al.\ 2016, JKAS, 49, 37
\bibitem[Kuang et al.(2022)]{Kuang2022}   Kuang, R., Zang, W., Jung Y. K., et al. 2022, \mnras, 516, 1704
\bibitem[Mao \& Paczy\'nski(1991)]{Mao1991} Mao, S., \& Paczy\'nski, B., 1991, \apj, 374, L37 
\bibitem[Nataf et al.(2013)]{Nataf2013} Nataf, D.~M., Gould, A., Fouqu\'e, P., et al.\ 2013, \apj, 769, 88
\bibitem[Paczy\'nski(1986)]{Paczynski1986} Paczy\'nski, B. 1986, \apj, 304, 1 
\bibitem[Poleski et al.(2014)]{Poleski2014}   Poleski, R., Skowron, J., Udalski, A. et al. 2014, \apj, 795, 42
\bibitem[Shin et al.(2023)]{Shin2023}   Shin, I.-G., Yee, J. C., Zang, W., et al. 2023, \aj, 166, 104
\bibitem[Skowron et al.(2018)]{Skowron2018}  Skowron, J., Ryu, Y. -H., Hwang, K. -H., et al. 2018, Acta Astron., 68, 43
%\bibitem[Udalski et al.(2005)]{Udalski2005}  Udalski, A., Jaroszy\'nski, M., Paczy\'nski, B., et al. 2005, \apj, 628, L109
\bibitem[Udalski et al.(2015)]{Udalski2015}  Udalski, A., Szyma\'nski, M. K., Szyma\'nski, G., et al. 2015, Acta Astron., 65, 1
\bibitem[Wo\'zniak(2000)]{Wozniak2000}   Wo\'zniak, P. R. 2000, Acta Astron., 50, 42
\bibitem[Yang(2023)]{Yang2023} Yang, H. 2023, in preparation 
\bibitem[Yee et al.(2012)]{Yee2012}    Yee, J. C., Shvartzvald, Y., Gal-Yam, A., et al. 2012, \apj, 755, 102 
\bibitem[Yoo et al.(2004)]{Yoo2004}    Yoo, J., DePoy, D. L., Gal-Yam, A. et al. 2004, \apj, 603, 139
\bibitem[Zang et al.(2021)]{Zang2021}   Zang, W., Han, C., Kondo, I., et al. 2021, Research in Astronomy and Astrophysics, 21, 239 
% ***********

%====================================
\vspace*{\fill}
\end{thebibliography}
\end{document}